%% file: paper.tex
\documentclass[12pt,preprint]{aastex}

\shorttitle{{\it Spitzer} IRS Observations of the Galactic Center}
\shortauthors{Simpson et al.}

\begin{document}

\title{{\it Spitzer} IRS Observations of the Galactic Center: Shocked Gas 
in the Radio Arc Bubble}

\author{Janet P. Simpson\altaffilmark{1,2}, Sean W. J. Colgan\altaffilmark{1}, 
Angela S. Cotera\altaffilmark{2}, Edwin F. Erickson\altaffilmark{1}, 
David J. Hollenbach\altaffilmark{3}, Michael J. Kaufman\altaffilmark{4}, and
Robert H. Rubin\altaffilmark{1,5}}

\altaffiltext{1}{NASA Ames Research Center, Mail Stop 245-6, Moffett Field, CA 94035-1000, USA}
\altaffiltext{2}{SETI Institute, 515 N Whisman Road, Mountain View, CA 94043, USA}
\altaffiltext{3}{NASA Ames Research Center, Mail Stop 245-3, Moffett Field, CA 94035-1000, USA}
\altaffiltext{4}{Dept of Physics, San Jose State University, One Washington Square, San Jose, CA 95192, USA}
\altaffiltext{5}{Orion Enterprises}

\email{jsimpson@mail.arc.nasa.gov}

\begin{abstract}
We present {\it Spitzer} IRS spectra (R $\sim600$, 10 -- 38 \micron) of 38 positions in the 
Galactic Center (GC), all at the same Galactic longitude and 
spanning $\pm 0.3^\circ$ in latitude.
Our positions include the Arches Cluster, the Arched Filaments, regions near the Quintuplet Cluster, 
the ``Bubble'' lying along the same line-of-sight as the molecular cloud G0.11$-$0.11, 
and the diffuse interstellar gas along the line-of-sight at higher Galactic latitudes. 
From measurements of the [\ion{O}{4}], [\ion{Ne}{2}], [\ion{Ne}{3}],
[\ion{Si}{2}], [\ion{S}{3}], [\ion{S}{4}], [\ion{Fe}{2}], [\ion{Fe}{3}], 
and H$_2$ S(0), S(1), and S(2) lines
we determine the gas excitation and ionic abundance ratios.
The Ne/H and S/H abundance ratios are $\sim 1.6$ times that of the Orion Nebula. 
The main source of excitation is photoionization, with the Arches Cluster
ionizing the Arched Filaments and the Quintuplet Cluster ionizing the gas nearby 
and at lower Galactic latitudes including the far side of the Bubble.
In addition, strong shocks ionize gas to O$^{+3}$ and destroy dust grains,
releasing iron into the gas phase 
(Fe/H$ \sim 1.3 \times 10^{-6}$ in the Arched Filaments and 
Fe/H$ \sim 8.8 \times 10^{-6}$ in the Bubble).
The shock effects are particularly noticeable in the center of the Bubble, 
but O$^{+3}$ is present in all positions.
We suggest that the shocks are due to the winds from the Quintuplet Cluster Wolf-Rayet stars.
On the other hand, the H$_2$ line ratios can be explained with 
multi-component models of warm molecular gas in photodissociation regions  
without the need for H$_2$ production in shocks.
\end{abstract}

\keywords{
Galaxy: center ---
infrared: ISM ---
ISM: abundances --- 
ISM: bubbles ---
\ion{H}{2} regions
}

\section{Introduction}
The Milky Way's Galactic Center (GC) includes a quiescent, 
$\sim 4 \times 10^6$ M$_\odot$ black hole (Ghez et al. 2005; Eisenhauer et al. 2005), 
three clusters of young massive stars, and massive molecular clouds
(see the reviews in Sch\"odel et al. 2006).
Figure 1a shows a radio image of the region from Yusef-Zadeh \& Morris (1987a), 
where the labels refer to the objects discussed in this paper.
The Radio Arc (Yusef-Zadeh et al. 1984) consists of non-thermally emitting 
linear filaments perpendicular to the Galactic plane; 
the Arched Filaments and Sickle, on the other hand, 
are thermal emitters, as is seen from their production of radio recombination lines 
(e.g., Yusef-Zadeh \& Morris 1987b; Morris \& Yusef-Zadeh 1989; Lang et al. 1997, 2001).
Figure 1b shows a Band E (21 \micron) image from the {\it Midcourse Space Experiment, MSX} 
(Egan et al. 1998; Price et al. 2001).
At 21 \micron, most of the emission comes from warm dust.

Sgr A West is an \ion{H}{2} region at the very center of the Galaxy,
to which we assume a distance of 8~kpc. 
It contains a cluster of massive stars and the black hole,
which is coincident with radio source Sgr A*.
The other two clusters of massive stars, the Arches Cluster and the Quintuplet Cluster,
are located about 25 pc away in the plane of the sky.
The Arches Cluster, whose O and Wolf-Rayet (WR) stars were 
first discussed by Nagata et al. (1995) and Cotera et al. (1996),
contains $\sim 10^4$~M$_\odot$ (Stolte et al. 2002),
emits $\sim 10^{51.6}$ ionizing photons s$^{-1}$ (Figer et al. 2002),
and is suggested to have an age of $\sim 2.5$ Myr (Figer et al. 2002).
The Quintuplet Cluster, named after the five luminous near-infrared (NIR) objects
that may be dusty carbon WR stars (see the discussion by Moneti et al. 2001),
is a little smaller ($10^{50.9}$ s$^{-1}$ ionizing photons, Figer et al. 1999b) 
than the Arches Cluster and probably older (Figer et al. 1999a, 1999b).

Given that it takes $\lesssim 1$ Myr for a cluster to orbit the GC,
it is probable that either or both the Arches and Quintuplet Clusters 
are now separated from their natal molecular clouds.
Any interactions with the GC's massive molecular clouds come 
from the long range effects of the clusters' high ionizing luminosities 
and the strong winds from their massive O and WR stars.
With discovery of the hot stars of the Arches and Quintuplet Clusters,
it was immediately suggested that these clusters provide the ionizing photons 
needed to power the Arched Filaments and Sickle, respectively 
(e.g., Cotera et al. 1996; Figer et al. 1995; Timmermann et al. 1996).

Since the thermal ionized gas was discovered before the star clusters,
there were originally questions regarding the source of excitation of the gas
(e.g., Erickson et al. 1991).
With the discovery of the two clusters, proof was sought that these are 
the exciting clusters by measuring the excitation variations 
(i.e., the ionization structure)
as a function of distance from the clusters.
To do this, far infrared (FIR) lines were observed 
using the {\it Kuiper Airborne Observatory} ({\it KAO}) (Colgan et al. 1996; Simpson et al. 1997)
and the {\it Infrared Space Observatory} ({\it ISO})
(Rodr\'{i}guez-Fern\'andez et al. 2001b; Cotera et al. 2005).
For both the Arched Filaments (Colgan et al. 1996; Cotera et al. 2005)
and the Sickle (Simpson et al. 1997; Rodr\'{i}guez-Fern\'andez et al. 2001b),
the excitation decreases with distance from the Arches Cluster and 
Quintuplet Cluster, respectively, as expected for photoionization. 

However, the radial velocity structure of both gas and stars in this region is complex, 
and there could be substantial line-of-sight distances between the star clusters 
and the ionized gas.
The Arched Filaments have radial velocities ranging from $\sim -25$ km s$^{-1}$
at the northern end to $\sim -40$ km s$^{-1}$ closest to Sgr A 
with occasional $V_{\rm LSR} \sim 0$ km s$^{-1}$ at intermediate locations (Lang et al. 2001).
At many locations the observed lines show more than one velocity component.
Because negative radial velocities are contrary to the direction of Galactic rotation,
Lang et al. (2001) suggested that the gas is moving on an elongated orbit 
in the Galaxy's barred spiral potential.
The gas of the Sickle has an even larger range of velocities with 
$V_{\rm LSR}$ ranging from $\sim -40$ km s$^{-1}$ to $\sim 140$ km s$^{-1}$
(Yusef-Zadeh et al. 1997; Lang et al. 1997).
Unlike the Arched Filaments, the average $V_{\rm LSR}$ of the Sickle is positive,
$\sim 35$ km s$^{-1}$ (Lang et al. 1997).
In both the Arched Filament and the Sickle regions, there exist molecular clouds 
with velocities similar to those of the ionized gas 
(e.g., Serabyn \& G\"usten 1987, 1991; Lang et al. 2002).
The ionized gas is most likely the photoionized edges of these clouds
(Lang et al. 2001).

Even though it seems likely that the thermal gas in Fig.~1a is the photoionized edges 
of molecular clouds, the morphology seen in Fig.~1 suggests
that the cloud structures are not randomly placed in the GC.
In particular, the systems of semi-circular arcs suggest 
that the morphology has been influenced by stellar winds 
or supernova explosions.
The most notable pair of arcs has become known as
the ``Bubble'' (Levine et al. 1999) 
or the ``Radio Arc Bubble'' (Rodr\'{i}guez-Fern\'andez et al. 2001b). 
This is the circular open structure south of the Quintuplet Cluster 
whose rim is faint at radio frequencies (Fig.~1a) but prominent in  
the 21 \micron\ continuum produced by warm dust (Fig.~1b, Egan et al. 1998; 
Price et al. 2001). 

At the same position on the plane of the sky as the Bubble, 
there is a dense molecular cloud called either G0.11$-$0.11 (Tsuboi et al. 1997) 
or G0.13$-$0.13 (Oka et al. 2001).
The column densities through the cloud are so high that the cloud 
is optically thick even at mid-infrared (MIR) wavelengths ($A_V \sim 640 - 740$, Handa et al. 2006).
In fact, the G0.11$-$0.11 cloud is so optically thick that 
Handa et al. (2006) suggest that the dark area of the center of the Bubble
is the shadow of the cloud.
In this scenario the Bubble rim would be the illuminated exterior of the G0.11$-$0.11 cloud.
We note, though, that there is no hint of such an optically thick cloud 
in the {\it Spitzer Space Telescope} IRAC images of the region (Stolovy et al. 2006, 2007), 
at which wavelengths (3 -- 8 \micron) the cloud would be even more optically thick.
We question whether the Bubble is in fact related to the G0.11$-$0.11 cloud.
We suspect the Bubble is a low density, interstellar bubble of hot shocked gas,
lying between the Earth and the optically thick G0.11$-$0.11 cloud in the GC.

In this paper we present {\it Spitzer Space Telescope} observations 
(Program 3295, all AORs) 
of MIR spectra of 38 positions along a line 
approximately perpendicular to the Galactic plane in the GC.
The observed line of positions is shown in Figure 1.
The emission from the 2 -- 3 positions at each end of the line 
probably originate in the diffuse interstellar medium (ISM)
along the line-of-sight to the GC as well as in the GC itself.
Our goals are to clarify the source(s) of excitation of the Arched Filaments 
and the other thermal arcs 
and to determine the relationship of the Bubble to the clusters and
other filament structures.
We defer discussion of the continuum features seen in the spectra 
and the spectra of the Arches Cluster stars to later papers.
In \S 2 we present the observations; in \S 3 we describe the results,
including the gas excitation and atomic abundances;
in \S 4 we present the evidence for high velocity shocks in the Bubble;
in \S 5 we discuss the origins of the shocked gas in the Bubble;
and in \S 6 we summarize our conclusions.

\section{Observations}
We observed the Galactic Center region with the Infrared Spectrograph (IRS)
(Houck et al. 2004) on {\it Spitzer Space Telescope} (Werner et al. 2004)
with the Short-High (SH, 10 -- 19.5 \micron) and Long-High (LH, 19.5 -- 38 \micron) modules   
on 21 -- 23 March 2005.
The LH spectra were taken in the IRS Stare mode, where a position on the sky is
observed for a number of integrations (``cycles''), in this case four,
and then displaced by a third the distance along the slit for 
a second group of integration cycles.
Since the SH slit is half the size of the LH slit 
($4.7'' \times 11.3''$ vs. $11.1'' \times 22.3''$, respectively), 
the same position was mapped with the SH module 
in a 2 by 3 position map so that essentially the same area on the sky was covered.
As with LH, four or eight integrations (three for the Arches Cluster map positions) 
were taken at each telescope pointing to provide redundancy against cosmic ray hits.
The Arches Cluster was mapped over a $\sim 45''$ square;
however, for this paper we summed all the map positions so that the spectra are
dominated by the interstellar gas and dust and not by the cluster stars
(the most massive stars of the Arches Cluster are largely found in the central $10''$ core,
and $> 80$\% in the central $20''$, Stolte et al. 2002).
For each telescope pointing, the total integration times ranged from 
18 -- 120~s for SH and 24 -- 56~s for LH.
The positions and integration times were chosen such that the bright lines would not 
saturate the sensitive IRS arrays;
as a result, there is no spatial agreement with positions observed with any other IR telescope,
such as the {\it KAO} or {\it ISO}.
The coordinates of each of the 38 positions are given in Table~1
and are plotted in Fig.~1.

The data were reduced and calibrated with the S13.2 pipeline at the 
Spitzer Science Center (SSC).
The basic calibrated data (bcd images) for each telescope pointing 
were median-combined 
and cleaned of rogue pixels\footnote{http://ssc.spitzer.caltech.edu/irs/roguepixels}
 and noisy order edges (the ends of the slit).
The spectra were extracted using both the 
Spectroscopic Modeling, Analysis and Reduction Tool (SMART, Higdon et al. 2004) 
and the SSC extraction routines in the {\it Spitzer} IRS Custom Extraction tool 
(SPICE\footnote{http://ssc.spitzer.caltech.edu/postbcd/spice.html});
we use the SMART-extracted spectra in this paper 
(the results from both spectrum extraction tools are similar).
The LH spectra were defringed with SMART and 
IRSFRINGE\footnote{http://ssc.spitzer.caltech.edu/archanaly/contributed/irsfringe}.
Because the IRS was calibrated using point sources but all our sources (positions)
are extended, a correction for the telescope diffraction pattern as a function
of wavelength is necessary; these ``slitloss'' correction factors were taken
from tables provided with SPICE.

\subsection{Spectra and Line Fluxes}

Three examples of the observed spectra are shown in Figure~2
(corrections for offsets between the orders will be discussed by 
Simpson et al., in preparation); 
these positions are of the center of the Bubble (Fig.~2a), 
the W Arched Filament (Fig.~2b), and the position at the highest Galactic latitude (Fig.~2c).
The three spectra are very representative of GC spectra 
of the Bubble, \ion{H}{2} regions, and the diffuse ISM, respectively.

Fluxes for the faintest (noisiest) and brightest lines were estimated by integrating 
over the line profiles (e.g., Colgan et al. 1996).
These integrated fluxes are probably more accurate than the Gaussian fits where 
the line/continuum ratio $> 1$
since they require no assumptions about the shape of the line profile; 
the differences with Gaussian fits are less than $\sim 1 - 2$\% for these bright lines: 
[\ion{Ne}{2}], [\ion{Ne}{3}], [\ion{S}{3}], and [\ion{Si}{2}].
Line and continuum fluxes of all other lines were measured by fitting
Gaussian profiles plus a linear continuum to each of the line profiles
in the spectra using both IDL's CURVEFIT procedure and the procedures in 
SMART (Higdon et al. 2004).
The Gaussian fits should be more accurate when the line/continuum 
ratio $< 1$, since the continuum is fit simultaneously with the line.
Errors for all fluxes were estimated 
by computing the root-mean-squared deviation of the data from the fit;
because of the systematic effects of low-level, uncorrectable rogue pixels on the spectra,
these errors are consistently larger than errors estimated any other way,
such as the statistics of median averaging or the counting statistics plus read noise
from the SSC's pipeline.
The lines discussed in this paper are listed in Table~2,
along with references to the atomic or molecular physics parameters 
(wavelengths, transition probabilities, collisional cross sections) 
needed for further analysis.
The measured line fluxes are given in a machine-readable Table 3 
and selected continuum fluxes (wavelengths containing no PAH emission features)
are given in a machine-readable Table 4, available in the on-line edition. 

\subsection{Extinction}
It is well known that the GC has $\sim 30$ mag of visual extinction 
(see, e.g., Cotera et al. 2000);
this corresponds to an optical depth at 9.6 \micron\ of $\tau_{9.6} \sim 3$
(Roche \& Aitken 1985; Simpson et al. 1997).
However, it is clear from looking at our extracted spectra and various line ratios
(such as the [\ion{S}{3}] 18.7/33.5 \micron\ ratio) that
the extinction is not constant for all positions.
Unfortunately, because the spectra do not extend to wavelengths shorter than 10 \micron,
we could not estimate $\tau_{9.6}$ by fitting extinction models to the data
(e.g., Simpson \& Rubin 1990).
Instead, we estimate the extinction correction by assuming that all spectra
have essentially the same intrinsic shape given by the ratio 
of the 10 and 14 \micron\ continuum fluxes (where there are no strong PAH features) 
and fitting that ratio with a template of the MIR fluxes measured in the Orion Nebula 
(Simpson et al. 1998) in this 10 -- 14 \micron\ spectral region.
We will discuss this in more detail in our paper on the continuum features 
(Simpson et al., in preparation).
The law for the extinction versus wavelength for the GC of Chiar \& Tielens (2006) was used 
to correct line fluxes at longer wavelengths (see Appendix A).
The extinction coefficients for each line as a fraction of the coefficient at 9.6 \micron\ 
are given in Table~2.

\section{Results}
As can be seen in Figure 3, the strengths of the ionized lines 
correlate well with the Arches and Bubble structures
seen in the radio (e.g., Fig. 1) and MIR continua (Egan et al. 1998; Price et al. 2001).
This is especially true of the lines that are strong in \ion{H}{2} regions 
([\ion{Ne}{2}] 12.8 \micron, [\ion{Ne}{3}] 15.6 \micron, [\ion{S}{3}] 18.7 and 33.5 \micron, 
the [\ion{S}{3}] lines not being plotted 
because they are very similar to the [\ion{Ne}{2}] line). 
Lines prominent in photodissociation regions (PDRs)  
([\ion{Si}{2}] 34.8 \micron\ and [\ion{Fe}{2}] 26.0 \micron)
also peak in the Arched Filaments and the Bubble rim, 
but there is also some contribution from the diffuse ionized gas
along the line-of-sight to the GC.
The 2 -- 3 positions at the ends of the line almost certainly sample
substantial amounts of foreground gas;
however, we do not subtract these line fluxes as ``background''
because (1) this diffuse ionized gas is interesting in itself 
and (2) the half a degree in Galactic latitude with the Galactic plane
in the middle almost certainly precludes the ``background'' from 
being uniform or flat.

\subsection{Radial Velocities}
Even though the resolution of the IRS SH and LH modules is only $\sim 500$ km s$^{-1}$,
as a result of our high signal/noise ratios and the IRS stability 
the observed radial velocities
of the strongest lines have uncertainties as small as 10 -- 20 km s$^{-1}$. 
These uncertainties are estimated from the dispersion of the differences in velocities of 
two lines that should have the same velocity at the same position, 
such as the two [\ion{S}{3}] lines.
These velocities are plotted in Figure~4.
Because the wavelength calibration of the IRS is still incomplete, 
the velocities of the ionized lines were normalized
to the radial velocity observed at Position 31 in the E2 Arched Filament 
by Lang et al. (2001, $V_{\rm LSR} = -15$ km s$^{-1}$). 
The dotted line is the average velocity of the four strongest ionized lines
([\ion{Si}{2}], [\ion{Ne}{2}], and [\ion{S}{3}]); these velocities are listed in Table~1.

The velocities of the H$_2$ lines were normalized as follows:
The CO J = 4 -- 3 intensities as a function of velocity and Galactic latitude 
were extracted from the data cube of CO measurements of Martin et al. (2004)
at Galactic longitude 0.121\degr.
Intensity-weighted average velocities were calculated for each latitude; 
these are plotted as a solid line in Fig.~4.
Our H$_2$ velocities were normalized by requiring that the average of all 
our H$_2$ velocities be equal to the average of these CO velocities  
sampled at the same Galactic latitudes as our data.

The excellent agreement of the strong, ionized lines 
shows that {\it Spitzer} can be used to 
measure relative radial velocities.
Velocity gradients of as much as 5 -- 9 km s$^{-1}$ arcsec$^{-1}$ can even be seen  
in the [\ion{Ne}{2}] and [\ion{S}{3}] lines in a few of the SH 2 by 3 maps.
The velocity of the weaker [\ion{Ne}{3}] line is not as consistent --- the line centroid 
is less well determined and the line may include 
a larger percentage of high-excitation diffuse ISM gas at a different velocity.

In fact, there are indications of multiple velocity components 
in Bubble Positions 15, 16, 19, 20, 21, and Bubble Rim Position 23.
Not only are these the positions with velocity gradients in the SH ionized lines,
the individual spectra for these positions have line widths 
significantly (25 -- 90 km s$^{-1}$) larger 
than the instrumental line width, with Position 19 being the most extreme
(leading to deconvolved full widths at half maximum $> 250$ km s$^{-1}$).
Most likely, Bubble Position 18,  which has one of the highest 
positive velocities in Fig. 4, is missing the negative velocity components 
of the other Bubble positions and thus has relatively narrow lines.
In general, the brighter lines are dominated by strong, single-velocity components 
and thus do not show any significant line broadening over the instrumental width.

The observed radial velocity pattern suggests that 
the Arched Filaments and Bubble rim structure have a substantial line-of-sight component
to their distances.
The average velocity of the ionized lines (dotted in Fig.~4) 
shows significant structure on scales $\lesssim 2$ pc,
whereas the $V_{\rm LSR}$ profile in the Arched Filaments is quite smooth.
If the negative-velocity gas is in front of (behind)  
the positive-velocity gas, the negative-velocity Arched Filaments are 
closer to (farther from) the Sun
than the positive-velocity, southern rim of the Bubble (Positions 10, 11, and 12).
Position 20, with its large negative velocity,  
may be part of the northern rim of the Bubble. 
Some of these Bubble positions have broad lines, indicative of 
an average over both positive and negative velocity components. 
The fact that the Bubble positions (14 -- 21) sometimes have broad lines 
and sometimes do not (but instead have large absolute velocities) 
is probably due to the ionized gas being clumpy. 
Positions 22 and 25 may be the southern extension of the Sickle
(``Sickle Handle'' in Table~1), 
since they have positive velocity like the Sickle (e.g., Lang et al. 1997; 
Simpson et al. 1997).
If the apparent Filaments and Bubble structures are caused by a wind or explosion 
some time in the past, one would expect the foreground components to have
negative velocities due to expansion. 
It would be interesting to see a complete map at higher spectral resolution
of the velocities in the Bubble region 
since surely, the structure seen here is affected by the sample locations.

\subsection{Molecular Hydrogen and PDRs}

Unlike the ionized lines, the H$_2$ line fluxes peak (Fig. 3a) near the Galactic plane 
(latitude $\sim -0.05^\circ$) and have a relatively flat intensity distribution
with Galactic latitude,  
similar to the distribution seen in warm CO J = 4 -- 3 (Martin et al. 2004).
The radial velocities of the H$_2$ lines also show a distinctly different 
pattern from those of the ionized lines (Fig.~4) (correlation coefficients $< 0$).
This suggests that the warm molecular hydrogen  
is not associated with the same Arched Filament and Bubble structures 
that are seen in the atomic lines and dust continuum.
The H$_2$ lines probably arise from numerous PDRs along the line-of-sight to the GC
as well as molecular clouds in the GC itself (see the Appendix).
We will leave additional discussion of 
the PDR properties of the molecular clouds to a later paper. 

However, we do note that the [\ion{Si}{2}] lines 
(and the [\ion{Fe}{2}] lines, although with poorer signal/noise)  
show the same velocity structure  (Fig.~4) as the ionized lines
(correlation coefficients $\gtrsim 0.8$), 
in addition to showing a similar intensity pattern (Fig.~3)
(correlation coefficients $\gtrsim 0.7$).
Hence, we ascribe these low-excitation ionized lines 
to the photoionized gas of the GC,
rather than to the largely neutral gas associated with PDRs.

\subsection{Electron Temperatures}

Computations of \ion{H}{2} region abundances require knowing the 
electron temperature, $T_e$.
Although the infrared forbidden lines have low sensitivity to $T_e$ 
(the forbidden line emissivities range from being proportional 
to $T_e^{-0.25}$ to $T_e^{-0.45}$),
recombination lines are quite sensitive to $T_e$ --- 
the emissivity of the H 7 -- 6 line is proportional to $T_e^{-1.3}$.
Measurements of $T_e$ have been published for the Arched Filaments 
by Lang et al. (2001).
A typical value for our Arched Filament positions is $T_e \sim 6000$~K.
We adopt this value for all positions and include an estimated uncertainty of 500~K.
In fact, we expect variations in $T_e$ and will discuss the effects 
of other values of $T_e$ in later sections. 

\subsection{Electron Densities and Corrections to the Estimated Extinction}

Electron densities, $N_e$, were estimated from the flux ratios of 
the density-sensitive [\ion{S}{3}] 18.7 and 33.5 \micron\ lines 
(e.g., Rubin 1989; Simpson et al. 1995).
The [\ion{S}{3}] 18.7 to 33.5 \micron\ flux ratio is also sensitive to extinction.
For 8 positions the corrected 18.7/33.5 \micron\ line ratio (\S 2.2) 
is below that produced by even the lowest possible $N_e$,
suggesting that an additional extinction correction is necessary.
For these positions the estimated $\tau_{9.6}$ was increased by 0.1 -- 0.9 until 
the [\ion{S}{3}] 18.7/33.5 \micron\ line ratio corresponded to a minimum $N_e \sim 10$ cm$^{-3}$.
The estimated values of $\tau_{9.6}$ are listed in Table~1.
Since both [\ion{S}{3}] lines are very strong, the chief uncertainty for $N_e$ is 
due to uncertainties in the extinction correction. 
In order to include at least some of this extinction correction uncertainty 
in the error analysis, 
we assumed that the one sigma uncertainty in the optical depth is 10\% 
of the value of $\tau_{9.6}$ in Table~1, 
multiplied by the factor appropriate for the wavelength of each line 
($\tau_\lambda/\tau_{9.6}$) given in Table~2, 
and propagated this uncertainty through all succeeding ionic ratio calculations. 
The computed densities are plotted in Figure 5.
The highest density is at Position 31, the brightest observed knot in the Arched Filaments.

Table 1 also contains the emission measure ($= \int N_e^2 dl$) needed to 
produce the observed H 7 -- 6 recombination lines.
These emission measures when combined with the minimum $N_e \sim 10$ cm$^{-3}$ 
require path lengths of ionized gas $\sim 100 - 500$ pc.
Such long path lengths are not unreasonable 
for the three end positions (1, 2, and 3) at negative Galactic latitudes 
but they may be an indication that at some positions, especially Positions 14 and 15 in the Bubble, 
the minimum density, and hence the extinction, should be larger. 
For most of the rest of the positions, where $N_e \gtrsim 50$ cm$^{-3}$, 
the pathlengths are only a few pc.
From this we infer that the gas emitting the ionized lines is clumpy, 
both within the Bubble and especially in the Arched Filaments.

\subsection{Ionic and Total Abundance Ratios}

Using the extinction-corrected fluxes and $N_e$ shown in Fig.~5, 
we computed various ionic ratios.
The ratios (Ne$^+$ + Ne$^{++}$)/H$^+$ and (S$^{++}$ + S$^{+3}$)/H$^+$
are plotted in Figure~6.
Although we would expect the Ne/H and S/H ratios to be constant,
we see that neither ratio is constant for all Galactic latitudes.
The usual explanation for variable abundance ratios in a single source 
is that some ionization state is not being counted --- this is certainly 
the case for the S/H ratio, as estimated by the (S$^{++}$ + S$^{+3}$)/H$^+$ ratio,
where there could be significant unmeasured S$^+$.
On the other hand, for \ion{H}{2} regions 
we expect $\gtrsim 99$\% of the neon to be either singly or doubly ionized, 
and even in partially-ionized diffuse ISM we expect neon to be essentially 
as ionized as hydrogen.
The sizable error bars are due almost entirely to the measurements of
the relatively weak H 7 -- 6 line; 
the 500~K uncertainty in $T_e$ adds an additional $\sim 10$\% to the error bars.
We suggest that the reason for the apparent variations in the Ne/H abundances 
is our assumption of a constant $T_e = 6000$~K.
If we assumed much lower $T_e$ in the diffuse ISM for the positions away from 
the Arched Filaments, we would derive 
a much more constant (Ne$^+$ + Ne$^{++}$)/H$^+$ ratio.
For example, decreasing the assumed $T_e$ from 6000~K to 3000~K would increase 
the derived (Ne$^+$ + Ne$^{++}$)/H$^+$ abundance ratio by a factor of $\sim 2.1$ 
(and the (S$^{++}$ + S$^{+3}$)/H$^+$ abundance ratio by a factor of $\sim 1.9$).
Since the extinction coefficients for the H 7 -- 6 and [\ion{Ne}{2}] lines are almost identical 
(Table~2), the Ne$^+$/H$^+$ ratio is not sensitive to extinction uncertainties.

Because of the uncertainty in $T_e$ and 
the need for correction for S$^+$ in the diffuse ISM positions, 
we compute total abundances only for the Arched Filament positions.
These positions are chosen because they should require the least correction for S$^+$.
Here the Si$^+$/(Ne$^+$ + Ne$^{++}$) ratio is particularly low (Fig.~6c), 
from which we infer that silicon is mostly doubly ionized or more, 
as should be sulfur.
For Positions 29 -- 34 we find the average Ne/H$ = 1.63 \pm 0.08 \times 10^{-4}$,
the average S/H$ = 1.16 \pm 0.06 \times 10^{-5}$, and Ne/S$ = 13.6 \pm 1.1$.
These values of Ne/H and S/H are both about $1.6 \pm 0.1$ times that of the Orion Nebula
(Simpson et al. 2004, 1998). 
We note that with updated cross sections (Table 2; Chiar \& Tielens 2006) 
and extinction (Cotera et al. 2000), 
the S$^{++}$/H$^+$ ratio measured by Simpson et al. (1995) for G0.095+0.012, 
which is part of the Arched Filaments, is now $1.8 \pm 0.3 \times 10^{-5}$,
such that S/H$ \sim 2.0 \times 10^{-5}$, after correction for S$^+$ and S$^{+3}$.

Simpson \& Rubin (1990), Simpson et al. (1995), Afflerbach et al. (1997), 
and Mart\'{\i}n-Hern\'{a}ndez et al. (2002)
showed that the gradients of abundances with respect to hydrogen 
extend all the way into the GC. 
Their observed gradients for Ne/H and S/H ranged from $-$0.039 to $-$0.086 dex kpc$^{-1}$,
with a weighted average of $-$0.059 dex kpc$^{-1}$.
However, the Orion Nebula has a Galactocentric distance of $\sim 8.4$ kpc, 
thus making our observed Arched Filament/Orion Nebula ratio of 1.6 correspond to a gradient 
of only $-0.025$ dex kpc$^{-1}$.
Simpson et al. (1995) suggested that the gradient is flatter in the inner Galaxy 
due to the influence of the inner Galaxy bar producing radial mixing.
On the other hand, there are variations in the chemical composition of the GC 
that may show that the gas is not well-mixed.
Rodr\'{\i}guez-Fern\'andez \&  Mart\'{\i}n-Pintado (2005) 
observed the \ion{H}{2} region lines
at a number of positions with the {\it ISO} spectrometers.
Their measured Ne/H and S/H ratios ($2 - 4 \times 10^{-4}$ and $8 - 30 \times 10^{-6}$,
respectively) are mostly larger than our measured ratios;
however, their values include large uncertainties due to extinction 
and to the different aperture sizes at the wavelengths of Br$\alpha$ 
and the Ne and S lines.

\subsection{Sources of Excitation}

The \ion{H}{2} region excitation can be estimated from the flux ratio
of a doubly ionized line to a singly ionized line of the same element
if the ionization potentials of both lines are greater than that of hydrogen (13.6 eV).
For {\it Spitzer}, the best line pair is 
the [\ion{Ne}{3}]~15.6~\micron\ and [\ion{Ne}{2}]~12.8~\micron\ lines, 
since both lines have similar extinction.
The excitation determined from the [\ion{S}{4}] 10.5 
and [\ion{S}{3}] 18.7 and 33.5 \micron\ lines
is less reliable because of the high extinction at 10 \micron\ in the GC.
The excitation ratios of Ne and S are plotted in Figure~7.

The excitation of an \ion{H}{2} region decreases with distance from the exciting stars,
both because there is absorption of the high energy photons by gas closer to the stars 
(resulting in smaller Str\"omgren spheres for the higher ionization species)
and because the radiation field is more dilute at larger distances.
The latter effect is seen where the \ion{H}{2} region consists of clumps of gas 
at varying distances from the star with no gas interior to absorb the high energy photons.
This case is usually described as having a low value of the ionization parameter $U$, 
where $U$ is defined as the ratio of the photon density and the electron density
and is proportional to the photoionization rate divided by the recombination rate.
The effect of a dilute radiation field is that the doubly and triply ionized ions 
tend to recombine to lower ionization states even if the ionizing stellar 
spectral energy distribution (SED) 
is hard enough that most ions are at least doubly ionized in ordinary, compact \ion{H}{2} regions.
Good examples are the ions of silicon and sulfur, 
which are normally at least doubly ionized in \ion{H}{2} regions
(the singly ionized states commonly occur in PDRs).
Noticeable amounts of the singly ionized states of these elements occur 
when log $U~\lesssim -2$, regardless of the exciting SED.
Consequently, there can be gas that is simultaneously high-excitation 
(containing doubly and triply ionized atoms) 
due to high-energy photons of the stellar SED (or shocks) 
but contains substantial numbers of singly-ionized atoms 
due to excessive recombination at low $U$.

We can test for changes in the excitation due to distance from the exciting stars  
by observations of singly ionized silicon. 
The Si$^+$/Ne ratio is shown in Fig.~6c but an even better test 
is the [\ion{Si}{2}] 34.8/[\ion{S}{3}] 33.5 \micron\ line ratio,
because it has little dependence on $N_e$, $T_e$, or extinction.
Simpson et al. (1997) discussed this effect for models with varying $U$,
which they compared to observations of the gas in the Sickle surrounding the Quintuplet Cluster.
Figure 8 shows the [\ion{Si}{2}] 34.8/[\ion{S}{3}] 33.5 \micron\ line ratio for the GC positions.
For models of compact, spherical \ion{H}{2} regions, 
this ratio ranges from $\sim 0.1 - 0.3$ (Simpson et al. 1997), 
and for observed \ion{H}{2} regions, from $\sim 0.1$ to 
$\sim 0.6$ (Simpson et al. 1997; M. R. Haas, private communication).
From Figure 8 we infer that for all the positions 
with the [\ion{Si}{2}] 34.8/[\ion{S}{3}] 33.5 \micron\ line ratio $> 0.7$, 
the gas lies at substantial distances from the exciting stars. 

We can use this apparent decrease in excitation with distance from a potential source
of ionizing photons to help identify the actual ionizing stars. 
Because the Arched Filament Positions 29 -- 34 
have [\ion{Si}{2}] 34.8/[\ion{S}{3}] 33.5 \micron\ line ratios (Fig.~8) 
 typical of compact \ion{H}{2} regions, 
they must be ionized by stars relatively nearby. 
These stars must be the Arches Cluster stars 
and not the much more distant Quintuplet Cluster stars.
The neon and sulfur excitation (Fig.~7a and 7b) 
decreases as one looks to higher Galactic latitude from the 
Arches Cluster as would be expected if these filaments are ionized by
the Arches Cluster and are physically more distant.
Because the gas at the position (28) of the Arches Cluster itself has lower excitation 
than the two positions at immediately higher Galactic latitudes,
we believe that the Arches Cluster lies some distance away from the gas
emitting the observed lines 
(see also Cotera et al. 2005 and Lang et al. 2001).
Since the gas at Positions 22 -- 26 has higher excitation yet (Fig. 7), 
we believe that this gas is excited by the Quintuplet Cluster and 
not the Arches Cluster. 
These positions also have radial velocities (\S 3.1) more in agreement 
with the Quintuplet Cluster and Sickle than with the Arched Filaments. 

We show now that 
the hot stars of the Quintuplet Cluster almost certainly ionize the rims of the Bubble.
A question that motivated this project was whether the southern side of the Bubble rim 
might be ionized by hot stars lying outside the G0.11$-$0.11 molecular cloud 
to the south.
We find no indication of hot stars at more southerly latitudes than the Bubble --- 
such stars would produce doubly ionized Si and probably Ne.
However, we observed that 
the [\ion{Si}{2}] 34.8/[\ion{S}{3}] 33.5 \micron\ line ratio is high in the rim, indicating 
that the exciting stars lie at some large distance,  
and the Ne$^{++}$/Ne$^+$ ratio steadily decreases 
with distance from the Quintuplet Cluster.
The region near the Sickle is probably also ionized by the Quintuplet Cluster 
with the possible exception of very low-excitation Position 20, 
the same position that has the anomalously negative radial velocity of Fig.~4.
Or it may be that Position 20 is also ionized by the Quintuplet Cluster but
has a large line-of-sight separation from it.

We have used the photoionization code CLOUDY (version 07.02.00, 
last described by Ferland et al. 1998) 
to estimate the ionization structure of 
clumps of gas offset from the Quintuplet Cluster at distances ranging from 0.01 to 25 pc.
We assumed empty space between the gas clumps and the exciting stars.
The models used the 36 kK supergiant atmosphere of Sternberg et al. (2003), 
abundances equal to two times those of the Orion Nebula (Simpson et al. 2004), 
$N_e = 316$ cm$^{-3}$, and $10^{50.9}$ ionizing photons per second.
The results for the integrated Ne$^{++}$/(Ne$^+$ + Ne$^{++}$) and 
S$^{+3}$/(S$^{++}$ + S$^{+3}$) ratios are plotted in Figure 9;
also plotted are the ratios for Positions 4 -- 26 where the abscissa is 
the distance of the position from the Quintuplet Cluster on the plane of the sky.
We see that the observed Ne$^{++}$/(Ne$^+$ + Ne$^{++}$) ratios (Fig.~9a) agree reasonably well 
with the models, the differences probably being due to the need for some additional 
line-of-sight contribution to the distance.
Although the S$^{+3}$/(S$^{++}$ + S$^{+3}$) ratios (Fig.~9b) also decrease with distance,
the model ratios decrease much more steeply than the observations.
The factor of $\sim 4$ discrepancy in the S$^{+3}$ fraction 
between the observations of positions in the Bubble rim and the models
cannot be due to an overcorrection for extinction, 
since the Bubble rim has typically $\tau_{9.6} \lesssim 2$, which produces 
a correction of $e^{2*0.777} = 4.66$ at 10.5 \micron. 

The temperature of the warm dust also decreases with distance from the Quintuplet Cluster,
as is shown in Figure 9c.
Calculations of the emission from interstellar dust grains are usually made as a function 
of $\chi$, 
the ratio of the impinging radiation field divided by the ``local interstellar radiation field''.
Li \& Draine (2001) and Draine \& Li (2007) calculated the emission spectrum 
of dust grains under conditions where $\chi$ ranged from 0.3 to $10^4$.
They find that the spectrum $\lesssim 25$ \micron\ shows little variation with changes 
in the radiation field intensity until $\chi > 10^3$,
because for these wavelengths at the lower $\chi$, the emission 
mostly arises from single-photon heating of very small carbon grains. 
For $\chi \gtrsim 10^3$,
the 18.7/33.5 \micron\ continuum ratio increases substantially with increasing $\chi$ 
because there is now a contribution from larger, steady-state grains.
Since $\chi$ is observed to be $> 10^4$ in the Sickle (Simpson et al. 1997), 
it is reasonable to expect that the positions closest to the Quintuplet Cluster 
are in the regime of high enough radiation field intensity 
to affect the 18.7/33.5 \micron\ ratio.
The 18.7/33.5 \micron\ ratio appears to flatten out at large distances 
from the Quintuplet Cluster because $\chi$ is now $< 10^3$.

Rodr\'{\i}guez-Fern\'andez et al. (2001b) showed that the excitation of both the 
nitrogen and neon lines decreases with distance from the Quintuplet Cluster 
by comparing their {\it ISO} line measurements with CLOUDY models.
Most of their observations were of the Sickle region between the Quintuplet 
and Arches Clusters, but they did have five positions in the Bubble rim.
Our observations, made with the great sensitivity of {\it Spitzer},
show that the influence of the Quintuplet Cluster extends well beyond the 
inner edge of the Bubble rim
and includes heating the dust as well as ionizing the gas.

\section{High Velocity Shocks}

Figure 7 shows that the highest excitation occurs in the interior of the Bubble region
and in the diffuse ISM at the ends of the line of positions.
Surprisingly, the highest excitation does not occur close to either cluster --- 
either there is some additional source of excitation in the Bubble or 
the gas at the positions closest to the Quintuplet Cluster 
actually has a large line-of-sight distance to the cluster.
We suggest here that the source of the high excitation gas may be high velocity shocks
in the Bubble.

\subsection{Triply Ionized Oxygen}

Because photons that can doubly ionize helium ($> 54.4$ eV) 
are required for production of O$^{+3}$ ($IP = 54.9$ eV), 
triply ionized oxygen should not be produced in Galactic \ion{H}{2} regions
except those ionized by the most massive O stars. 
However,
the [\ion{O}{4}] 25.89 \micron\ line is detected at almost every position that we observed.
The exceptions (observed signal/noise $< 3$) occur, not where the line is weak,
but where the continuum, with its noise-producing rogue pixels, is very strong.
The normalized line profiles that show this line are plotted in Figure~10.
Fluxes in the LH beam (248 square arcsec) with $> 3 \sigma$ measurements 
range from $2.1 \times 10^{-17}$ W m$^{-2}$ to $1.5 \times 10^{-16}$ W m$^{-2}$.
The peak fluxes occur in Positions 19 -- 27 but the best signal/noise is
for positions in the Bubble and the diffuse ISM at high Galactic latitudes
where the line/continuum ratio is the largest.
We plot the O$^{+3}$/(Ne$^{+}$ + Ne$^{++}$) ratio in Figure~11.
Since the O/Ne ratio, which is expected to be approximately constant,
is observed to equal $\sim 4$ in Galactic \ion{H}{2} regions 
and O$^{+3}$/(Ne$^{+}$ + Ne$^{++}$) is $< 0.01$ here, 
we see that very little of the oxygen is O$^{+3}$ and 
therefore very little neon should be in higher ionization stages, too.

Although triply ionized oxygen should not be produced in Galactic \ion{H}{2} regions,
it has, in fact, been detected in some extra-galactic \ion{H}{2} regions.
However, either these are extremely low metallicity, 
which hardens the stellar ionizing SED,
or the exciting stars include hot WR stars (Schaerer \& Stasi\'nska 1999).
This is not the case in the GC --- we observe no enhancement 
of the O$^{+3}$/(Ne$^{+}$ + Ne$^{++}$) ratio at the position of the Arches Cluster, 
indicating that its WR stars are not effective at producing O$^{+3}$.

Lutz et al. (1998) discussed the [\ion{O}{4}] 
that they had observed with {\it ISO} in some galactic nuclei that are not low metallicity.
They attempted to reproduce the [\ion{Ne}{3}]/[\ion{Ne}{2}] 
and [\ion{O}{4}]/([\ion{Ne}{2}] + 0.44[\ion{Ne}{3}]) line ratios 
using CLOUDY (Ferland et al. 1998).
They modeled both a weak active galactic nucleus embedded in a starburst and 
an \ion{H}{2} region ionized by main-sequence stars plus a hot (80 kK) blackbody
to represent WR stars.
All their models showed the same pattern: 
in order to produce as much [\ion{O}{4}] emission as is observed, 
they had to include so much of the hot component 
that the [\ion{Ne}{3}]/[\ion{Ne}{2}] ratio was much larger than observed.
They noted that if there were small regions producing relatively strong [\ion{O}{4}] 
intermixed with low excitation gas, they might be able to produce the observed 
ratios; however, they considered it unlikely because they knew of no such regions
in the Galaxy or 30 Dor, and planetary nebulae produce too small a contribution
to be detected.
They suggested that the high excitation of [\ion{O}{4}] is caused by shocks.

We have used most recent version of CLOUDY (07.02.00) 
to make similar models, employing both black bodies 
and the supergiant SEDs of Sternberg et al. (2003),
who used the WM-BASIC model atmosphere code of Pauldrach et al. (2001).
Supergiant models are defined here as the atmosphere models with the smallest gravity, $g$, 
for each stellar effective temperature, $T_{\rm eff}$.
Supergiant SEDs are used because \ion{H}{2} region models computed with these SEDs as input 
give substantially better agreement with observed Ne$^{++}$ lines than 
the WM-BASIC SEDs with other gravities or computed with other model atmosphere codes 
(Simpson et al. 2004; Rubin et al. 2007). 
We emphasize that even though 
the actual stars may be something other than supergiants, the integrated cluster SEDs 
are similar to the calculated supergiant model SEDs.
The CLOUDY models have as input density = 316 cm$^{-3}$; 
Sternberg et al. SEDs with $T_{\rm eff}$ = 32 kK to 50 kK; 
number of ionizing photons $N_{\rm Lyc} = 10^{49}$, $10^{50}$, and $10^{51}$ s$^{-1}$; 
and inner shell radius $R_i$= 0.01, 1.0, and 10 pc.
The blackbody models have $T_{\rm eff}$ = 30 kK to 45 kK and similar other parameters.
Many of the CLOUDY models with blackbody SEDs have additional blackbody components 
with $T_{\rm eff} = 10^5$ and/or $10^6$ K to represent hot WR stars or 
the pervasive X-ray flux of the GC. 

Figure 12 shows the fraction of O$^{+3}$ versus the fraction of Ne$^{++}$ 
for both the \ion{H}{2} region models and our observations.
The only models using Sternberg et al. (2003) atmospheres (open circles) 
with enough O$^{+3}$ to show up on this plot are the two models with $T_{\rm eff} = 50$ kK,  
$N_{\rm Lyc} = 10^{49}$ s$^{-1}$, and $R_i =$ 0.01 or 1.0 pc
(since star clusters containing multiple 50 kK stars are not likely in the Galaxy,  
no models with 50 kK atmospheres and larger $N_{\rm Lyc}$ were computed).
The asterisks are the blackbodies with additional hot components for WR stars and X-rays.
As was earlier shown by Lutz et al. (1998), all the data fall to the upper left 
(larger O$^{+3}$/Ne$^{++}$) with respect to all the models calculated with believable SEDs 
(asterisks and open circles).

Before leaving this topic we test whether there is {\it any} SED that could 
be used in an \ion{H}{2} region model that would reproduce the observed ionic ratios.
To do this we computed models with the Sternberg et al. (2003) SED for 36 kK, 
(which has a sharp drop-off at 54.4 eV), 
but we arbitrarily increased the flux for photon energies $E > 54.5$ eV ($> 4.013$ ryd).
The other inputs were $R_i = 10$ pc and log $N_{\rm Lyc} = 50.5$.
Such SEDs do not affect ionic ratios of low-excitation ions, 
unlike SEDs where the flux at 54.5 eV is increased by the addition of a blackbody, 
which has substantial photon emission at energies lower than its maximum.
The ionic ratios from these models are plotted as open triangles in Fig. 12.
The lowest triangle on this plot is from the model 
that has a jump in the SED of $\sim 4$ orders of magnitude  
going from 54 eV to 54.5 eV.
In this model the SED is increased by 
a factor of $10^9$ over the original very low flux with $E > 54.5$ eV 
found in Sternberg et al's model atmospheres.
The open triangles connected by the solid line have an increase of a factor of $10^{10}$
over the original SED; their values of $R_i$ are 2, 5, 10, and 15 pc (top to bottom).
We see that the only model lying in the area of the observations is the model with 
$R_i = 10$ pc and a SED with a flux jump of a factor of $\sim 10^5$ at 54.5 eV.
SEDs with other flux jumps and $R_i$ could probably fit, but if the SED represents 
the ionizing source of {\it all} the positions, which all have {\it different} $R_i$,
it would not be possible for the same SED to fit all positions.
Moreover, it is difficult to imagine a physical process 
that would produce such strong emission in the He$^{++}$ continuum $> 54.5$ eV 
when the normal stellar SED has deep absorption.
We conclude that the O$^{+3}$ is not produced by photoionization.

The advantage of observing [\ion{O}{4}] in the GC is that we can locate 
the maximum of the high-excitation gas 
as being in a region that does {\it not} appear to contain the exciting stars.
We also can eliminate any possibility that the [\ion{O}{4}] is confined to 
small but high excitation regions as we detected it at all our positions.
According to the shock models of Contini \& Viegas (2001),
a shock velocity $\sim 100$ km s$^{-1}$ 
is adequate to produce the amount of [\ion{O}{4}]  that is observed.

\subsection{Gas-Phase Iron Production in Shocks}

Whereas O$^{+3}$ could be produced by high energy photons such as X-rays,
high gas-phase abundances of elements normally depleted onto grains 
indicate grain destruction in shocks.
Here we demonstrate that the gas-phase iron abundance is greatly increased 
in the center of the Bubble compared to the surrounding gas.
This is suggestive of shock production.

To compute the Fe abundance ratios we modeled the Fe$^+$ and Fe$^{++}$ ions 
with 36 and 34 level statistical equilibrium calculations, respectively 
(see Table~2 for atomic parameters and references).
Figure 10 shows the observed [\ion{Fe}{2}] lines and 
Figure 13 shows the derived (Fe$^{+}$~+~Fe$^{++}$)/(Ne$^{+}$~+~Ne$^{++}$) ratio.
Here we have plotted ratios computed using both the [\ion{Fe}{2}] 25.99 \micron\ line 
and the [\ion{Fe}{2}] 17.94 \micron\ line since the former includes 
substantial emission from the low temperature PDRs along the line-of-sight
(e.g., Kaufman et al. 2006) 
but the lower level of the latter is 1873 cm$^{-1}$ above ground and 
the line should arise only in the \ion{H}{2} region gas, 
like the [\ion{Fe}{3}], [\ion{Ne}{2}], and [\ion{Ne}{3}] lines.
Unfortunately, the [\ion{Fe}{2}] 17.94 \micron\ line is weak and is 
somewhat blended with the [\ion{P}{3}] 17.885 \micron\ line (Fig. 10a).
With the {\it Spitzer} resolution of $\sim 600$, the lines are separated 
by $\sim 1.7$ resolution elements and so can usually be distinguished;
however, the measurements have some additional uncertainty because 
the 2-line-Gaussian-fit routines in SMART would converge only if
the line width was fixed at the {\it Spitzer} IRS SH resolution.
There are two other [\ion{Fe}{2}] lines in the LH bandpass but they are either 
in a very noisy part of the wavelength range (35.35 \micron) or 
in a region with bad fringing (24.52 \micron).
The latter line is usually detectable, but since its signal/noise ratio is
much worse than the 17.94 \micron\ line, we do not use it.
The [\ion{Fe}{3}] 33.04 \micron\ line also is usually detectable (Fig. 2a),
but since it imparts no additional information with respect to $N_e$ or $T_e$
and its signal/noise ratio is so much poorer than the 22.9 \micron\ line that the 
33.0/22.9 line ratio is not usable for extinction estimates, 
we do not use it either.

We estimate the gas-phase abundances of iron in the GC by using our 
observed (Fe$^{+}$ + Fe$^{++}$)/(Ne$^{+}$ + Ne$^{++}$) ratios (Fig.~13): 
(Fe$^{+}$ + Fe$^{++}$)/(Ne$^{+}$ + Ne$^{++}$)$ = 0.0066 \pm 0.0016$ 
in the Arched Filaments and 
(Fe$^{+}$ + Fe$^{++}$)/(Ne$^{+}$ + Ne$^{++}$)$ = 0.043 \pm 0.004$ in the Bubble.
Since Fe$^{++}$ is moderately-low excitation, there can be substantial Fe$^{+3}$
in photoionized \ion{H}{2} regions
(see Rodr\'{\i}guez \& Rubin 2005 for a discussion of the uncertainties
in the measurements of Fe abundances in \ion{H}{2} regions).
We estimate the correction for Fe$^{+3}$ using both the CLOUDY models described above 
and models computed with the code NEBULA (Simpson et al. 2004; Rubin et al. 2007), 
noting that NEBULA and CLOUDY produce similar results for the 
same SEDs and other input parameters.
The ionization fractions computed with these models are plotted in Figure~14. 
As explained above, we prefer the models computed with the supergiant atmospheres 
of Sternberg et al. (2003), which are connected with the heavy solid line in the plot;
from these we estimate an ionization correction factor of $\sim 20$\%. 
This gives the gas-phase (Fe/Ne)$_{\rm gas}$ = 0.0082 and 0.054 in the Arched Filaments 
and Bubble, respectively.
Assuming that the Ne/H ratio observed in the Arched Filaments 
(Ne/H = $1.63 \times 10^{-4}$) is applicable to all the GC gas
and that Ne is not condensed onto grains,
by multiplying (Fe/Ne)$_{\rm gas}$ by Ne/H we estimate that 
(Fe/H)$_{\rm gas} \sim 1.3 \times 10^{-6}$ in the Arched Filaments and 
$ \sim 8.8 \times 10^{-6}$ in the Bubble.
There is an increase in the gas-phase abundance of Fe of a factor of $6.5 \pm 1.7$ 
in the Bubble with respect to the surrounding gas.

In order to estimate the depletion factors for Fe in the GC, one first 
needs to know the total Fe abundance in both grains and gas.
There are several ways to do this:

(1) There are interstellar absorption lines of O, Ne, and Fe from 13.4 -- 23.5 \AA\ 
that can be observed by {\it Chandra} and {\it XMM-Newton}.
The advantage of X-ray measurements is that the atoms in the grains produce 
sufficient absorption that total abundances are measured, 
not just gas phase abundances.
Using {\it Chandra},
Juett et al. (2006) observed absorption lines and edges in front of 7 neutron stars,
measuring averages of O/Ne $= 5.4 \pm 1.6$ and Fe/Ne $= 0.20 \pm 0.03$
(unfortunately, this soft X-ray wavelength range 
suffers too much interstellar extinction for observations of GC sources).
They derive the expression
\begin{equation}
({\rm Fe}/{\rm Ne})_{\rm ISM} = ({\rm Fe}/{\rm Ne})_{\rm meas} / (1 - \beta + f\beta)
\end{equation}
to correct for optical depth effects in the larger grains,
where (Fe/Ne)$_{\rm ISM}$ is the total gas plus grains ratio,
(Fe/Ne)$_{\rm meas}$ is the measured average ratio given above,
$\beta$ is the fraction of Fe residing in grains, 
and $f$ is the factor accounting for grain optical depths and the grain size distribution.
They derive $f = 0.88$ for Fe and $f = 0.90$ for O.
Using eq. (1) we find that (Fe/Ne)$_{\rm ISM}$ ranges from 0.20 to 0.23 and 
(O/Ne)$_{\rm ISM}$ ranges from 5.4 to 6.1 for $0 < \beta < 1$.

So that they would not need to make corrections for grains, Yao et al. (2006) measured 
high ionization absorption lines of Ne and Fe in the low-mass X-ray binary 4U 1820$-$303,
with the results that O/Ne = 6.2 and Fe/Ne = 0.30.

(2) There are X-ray emission lines of Fe at 6.4 and 6.7 keV. 
Wang et al. (2006) measured the spectra of the Arches Cluster stars with {\it Chandra}.
From the 6.7 keV line --- which probes highly ionized gas in the shocked cluster winds --- 
they estimated that the Fe abundance is 1.8 times Solar, referring to the Solar abundances 
of Anders \& Grevesse (1989), or Fe/H = $8.4 \times 10^{-5}$.
From measurements of the spectrum of the diffuse emission surrounding the Arches Cluster 
with {\it Suzaku},
Tsujimoto et al. (2007) estimated that the abundance of Fe is Solar, again with respect to 
the abundance summary of Anders \& Grevesse, such that Fe/H$ \sim 5 \times 10^{-5}$.
Koyama et al. (2007), using {\it Suzaku}, estimated an average Fe column 
density$ = 9.7 \times 10^{18}$ cm$^{-2}$ for the diffuse gas in the GC,
from which they calculated that the iron abundance is 
Fe/H$ \sim 1.6 \times 10^{-4}$, 
assuming that the column density of H is $6 \times 10^{22}$ cm$^{-2}$.
For these iron abundances, we estimate Fe/Ne$ \sim 0.5$, 0.3, or 1.0, respectively.

(3) By observing the spectra of cool stars in the GC, 
Ram\'{\i}rez et al. (2000) and Blum et al. (2003) found that 
the abundance of iron in the GC central cluster is $\sim$ Solar.
From the spectra of hot stars in the central cluster and the Quintuplet cluster,
Najarro (2006) also found a Solar abundance of iron.
These results are in contradiction to the computations giving a steep gradient 
for Fe in the Galaxy (e.g., Cescutti et al. 2007).
The most recent compilation of Solar abundances is that of Asplund et al. (2005), 
who give Fe/H$ = 2.8 \times 10^{-5}$.
From this we estimate that the total Fe/Ne$ \sim 0.17$.

In summary, estimates of the total Fe/Ne ratio in the GC range from $\sim 0.17$ to 
$\sim 1.0$. 
With these estimates (0.17/0.30/1.0) of the Fe/Ne ratio and our measurements, 
we find that the fraction of Fe in the gas phase is 
0.048/0.027/0.0082 and 0.32/0.18/0.054 for the Arched Filaments and 
the Bubble, respectively.
Thus, in spite of the factor of 6.5 increase of gas-phase iron in the Bubble with respect 
to the surrounding ISM,
most Fe is still incorporated in grains and is not in the gas-phase.

Dust grains can be destroyed in shocks, both by grains shattering 
into smaller particles and being vaporized in grain-grain collisions and
by removal of individual atoms by photons and by high-velocity gas
(see Jones 2004 for a review).
In high velocity shocks, $V_s > 200$ km s$^{-1}$, the grains are stopped in the hot gas.
Since in hot gas H$^+$ and He$^+$ have the same kinetic energies ($\sim {3 \over 2} kT$)
but H$^+$ is more abundant, sputtering by protons dominates. 
In slower shocks, however, most of the destruction is done by the betatron acceleration
of grains, which then move at high speeds ($\sim V_s$) through the gas.
For this case, the collision velocity is the same for H$^+$ and He$^{++}$,
and since the He ions have 4 times the energy of the protons, 
they dominate the sputtering, even though they are less abundant by a factor of ten.

Jones et al. (1994, 1996) computed grain destruction models for 
a variety of shock velocities, initial densities $n_0$, and magnetic fields $B_0$.
Both $n_0$ and $B_0$ in the GC are probably higher 
than those used in any of their models (especially $B_0$)
but the destruction rate should scale as $B_0 n_0^{-1/2}$.
The observed depletions in the Bubble correspond to shock velocities 
$\sim 100 - 50$ km s$^{-1}$, depending on which Fe gas-phase fraction is correct.
These velocities are plausible if the shocks are due to the expansion of the Bubble 
(see the range of velocities in Table~ 1).

We note that there is no obvious increase in the Si$^+$/H$^+$ ratio in the Bubble
compared to the Bubble exterior that cannot be attributed 
to normal \ion{H}{2} region photoionization, where most Si is Si$^{++}$ or higher.
In the Bubble the average Si$^+$/(Ne$^+ + $Ne$^{++}$)$ = 0.102 \pm 0.003$, 
from which we estimate that  Si$^+$/H$ \sim 1.7\times10^{-5}$.
This is about 0.5 times Solar (e.g., Asplund et al. 2005).
However, if one of the higher depletion estimates for Fe is correct, 
it would require only a small fraction of the grains to be destroyed in the Bubble,
producing a non-detectable increase in the gas-phase Si abundance.

\section{Discussion}

Bubbles around hot stars and star clusters are commonly observed 
in other galaxies (e.g., Oey \& Massey 1994; Oey 1996)
and in the Milky Way (Churchwell et al. 2006).
Early analytic models were made by Weaver et al. (1977),
who modeled the time-varying stellar wind of a main-sequence star 
expanding into the ISM.
Later, Garc\'{\i}a-Segura \& Mac Low (1995a, 1995b) made both analytic and 
numerical models of stellar winds from WR stars. 
When these winds encounter gas from previous mass-loss episodes 
(main-sequence and red super-giant stages) and/or interstellar gas, 
a shock results, producing high temperatures and hence high pressures.
A shell is formed surrounding the high-pressure interior; 
the expanding shell sweeps up the exterior interstellar gas, shocking and heating it.
The temperatures of the lower-density bubble interior are sufficiently high that,
at least sometimes, X-rays are produced (Chu \& Mac Low 1990).
Although these models do predict the formation of bubbles, 
Oey (1996) found that the predicted bubbles are much larger than observed.
Oey \& Garc\'{\i}a-Segura (2004) found that numerical hydrodynamic and radiative models 
that include substantial pressure, $P/k \sim 10^5$ cm$^{-3}$ K, in the ambient ISM 
result in bubbles and shells much more similar in size to the observed bubbles.

The Bubble seen in the GC is fairly small (radius $\sim 10$ pc) compared 
to the bubbles in the Large Magellanic Cloud observed by Oey (1996) and
Oey \& Garc\'{\i}a-Segura (2004).
On the other hand, the ambient ISM pressure in the GC is probably at least an order 
of magnitude larger --- for the GC as a whole, 
Spergel \& Blitz (1992) estimate $P/k \sim 5 \times 10^6$ cm$^{-3}$ K.
Rodr\'{\i}guez-Fern\'andez et al. (2001a) measured CO lines with the {\it IRAM} 
30 m telescope and H$_2$ lines with {\it ISO} 
at their position M+0.16$-$0.10 in the center of the Bubble.
They estimated an H$_2$ density of $\sim 10^4$ cm$^{-3}$ from their CO measurements 
and a temperature of $157 - 900$~K from their H$_2$ line measurements, 
giving $P/k = nT \sim$ few $\times 10^6$ cm$^{-3}$ K.
This gas is most likely found in a PDR, probably the edge of the 
molecular clouds in this region.
The turbulent pressure must be large, given that the H$_2$ line widths 
(180 km s$^{-1}$) are much larger than the sound speed.
Possibly the largest contributor to the ISM pressure is the magnetic field, $B$.
Yusef-Zadeh \& Morris (1987a) suggested that the magnetic field in
the linear non-thermal filaments of the Radio Arc is as high as 1~mG.
If so, the magnetic pressure ($P_B/k = 2.9 \times 10^8 B_{mG}^2$ cm$^{-3}$ K)
dominates the gas pressure, although it would be found in both the interior and
the exterior of the Bubble.
If the linear filaments do not thread through the Bubble,  
$B$ could be much lower.
LaRosa et al. (2005) found that the global field of the GC is only of order $10~\mu$G 
with an upper limit of $\lesssim 100~\mu$G.
Since the magnetic pressure is proportional to $B^2$,
the global magnetic pressure, in this case, would not contribute significantly to 
the ambient ISM pressure.

Looking at the models of Oey \& Garc\'{\i}a-Segura (2004),
we suspect that similar models with ambient ISM pressure $\gtrsim 10^6$ cm$^{-3}$ K 
could produce bubbles with radii of order 10 -- 20~pc, such as are observed 
in the high-pressure environment of the GC.
The star clusters producing the bubbles in their models have ionizing luminosities 
$N_{\rm Lyc} = 10^{49.4} - 10^{50.2}$ s$^{-1}$ 
(we include here only their ``Pre-SN'' clusters because 
these bubbles have thicker observed shells like the Bubble 
and not the very thin shells of ``Post-SN'' clusters).
The Quintuplet Cluster lies within the Bubble, although not at its center,
and has $N_{\rm Lyc} \sim 10^{50.9}$ photons s$^{-1}$ 
and  \.{M} $\gtrsim 7 \times 10^{-4}$ M$_\odot$ yr$^{-1}$ (Lang et al. 2005).
Since the sizes of observed expanding shells are always much smaller 
than what is expected from the kinetic energy of the stellar wind,
we speculate that the Bubble is produced by the Quintuplet Cluster. 
The Bubble asymmetry is probably caused by the asymmetry in the surrounding molecular clouds,
such that the Quintuplet Cluster is close to the cloud next to the Sickle
(Simpson et al. 1997).
Before the Cluster wind started creating the Bubble, 
there was a gradient in the gas extending from the 
$V_{\rm LSR} = +25$ km s$^{-1}$ G0.20$-$0.033 molecular cloud (Serabyn \& G\"usten 1991)
through the location of the stars to lower Galactic latitudes.

We can ask about the extent of the shocked gas in the Galactic longitude direction.
The {\it Spitzer} archive includes Program 18 (PI: J. Houck), which 
contains SH and LH spectra (all short integrations) of 18 FIR sources from 
Galactic longitudes $-$0.96 to +3.06. 
We searched these data for the [\ion{O}{4}] line and other indicators of 
high-velocity shocked gas.
There is one position in the middle of the Bubble with [\ion{O}{4}] at similar intensity 
to our own Bubble positions;
the two positions on either side, at Galactic longitude $-$0.15 and +0.24,
have [\ion{O}{4}] line fluxes factors of 4 -- 6 fainter.
The next positions on either side have no [\ion{O}{4}].
The [\ion{Fe}{3}] flux also peaks in their Bubble position.
The relatively small extent of the shocked gas in Galactic longitude is
another indication that the Bubble is local to the Quintuplet Cluster region.

Rodr\'{\i}guez-Fern\'andez \&  Mart\'{\i}n-Pintado (2005) observed 
most of these positions with {\it ISO}.
They also found the gas in the center of the Bubble (position M+0.16$-$0.10) 
has the highest excitation as seen by the [\ion{N}{3}]57/[\ion{N}{2}]122 \micron\ 
and [\ion{Ne}{3}] 15.6/[\ion{Ne}{2}] 12.8 \micron\ line ratios.
They conclude that the \ion{H}{2} region gas at this position  
is mainly excited by the Quintuplet Cluster
but the gas seen in lines-of-sight more distant from the Arches and Quintuplet Clusters 
must have their own ionizing sources 
since their ionization does not decrease with distance from the clusters.
This is in accord with our conclusion that 
the Bubble is excited by the Quintuplet Cluster stars.

\section{Summary and Conclusions} 

We present {\it Spitzer} IRS spectra of the Arched Filaments, Sickle Handle, 
and Bubble in the GC, along with spectra of the diffuse ISM 
at Galactic latitudes up to $\pm 0.3^\circ$ from the Galactic plane.
In this paper we concentrate on interpreting the ionic lines, using 
the H$_2$ lines and continuum mainly for estimation of the interstellar extinction.

From our estimates of the radial velocities of the H$_2$ lines, 
accurate to 10 -- 20 km s$^{-1}$,
we infer that there are significant contributions to the H$_2$ lines seen in our spectra 
from gas at locations along the line-of-sight other than the location of the ionized gas.
The H$_2$ line ratios are consistent with emission from multiple components 
at different temperatures. 
The low excitation ionized lines ([\ion{Si}{2}] and [\ion{Fe}{2}])
have both radial velocities and relative intensities better correlated with 
those of the more highly ionized \ion{H}{2} region lines 
than with the H$_2$ lines. 
We conclude that these lines are formed in both the \ion{H}{2} regions 
and in PDRs closely associated with these \ion{H}{2} regions.

The abundances of neon and sulfur with respect to hydrogen are a factor of $\sim 1.6$ 
larger than those of the Orion Nebula, the standard for interstellar abundance studies.
We note that these abundances are quite sensitive to $T_e$  
because of the sensitivity of the H 7 -- 6 line emissivity to $T_e$,
which we assume equals $6000 \pm 500$ K.
The effect is such that if one assumes a lower $T_e$, as though there were more cooling
from higher metallicity, 
one will as a result deduce higher metallicities. 
This shows the importance of accurate measurements of $T_e$.

The measurement positions that we chose sample both sides of various ionization structures,
thereby enabling identification of the direction from which the ionizing photons originate.
The south rim ($b \sim -0.17^\circ$  to $-0.25^\circ$) of the Bubble 
is ionized by a source at higher Galactic latitude, probably the Quintuplet Cluster.
This part of the Bubble rim is not the outer edge of the G0.11$-$0.11 molecular cloud,
which lies along the same line-of-sight.

The Arched Filaments are ionized by the Arches Cluster.
The Arched Filament positions are the only locations 
where the Si$^+$/H$^+$ abundance ratio is relatively low,
probably because silicon is largely doubly ionized there.
Such an ionization pattern can happen only when the ionizing stars are nearby, 
and the Arches Cluster is much closer to the Arched Filaments 
than any other known source that produces a substantial number of ionizing photons.

The highest excitation, as seen by the indicators Ne$^{++}$/Ne$^+$, 
S$^{+3}$/S$^{++}$, Si$^+$/S$^{++}$, and O$^{+3}$/(Ne$^{+}$~+~Ne$^{++}$), 
is found in the center of the Bubble and not at the locations 
of the Arches and Quintuplet Clusters.
At the present time there is probably a several pc line-of-sight offset
between the Arches Cluster and the ionized gas.
This is not unreasonable since the clusters are of ages such that they 
have orbited the Galaxy at least once since their formation.

There are multiple indicators of strong shocks,
peaking in the center of the Bubble.
This is indicated both by the presence of O$^{+3}$,
which cannot be photoionized by stars with as large a metallicity as exists in the GC,
and by the increase by a factor of 5 -- 8 of gas-phase iron
from Fe/H $\sim 1.3 \times 10^{-6}$ in the Arched Filaments 
to Fe/H $\sim 9 \times 10^{-6}$ in the Bubble. 
This increase in the gas-phase iron abundance is probably 
produced by shock destruction of grains. 
The strong shocks possibly indicate the location of a violent outflow 
in the recent past.
We suggest that the source of this outflow is the Quintuplet Cluster. 
A shock velocity $\lesssim 100$ km s$^{-1}$ is adequate to produce 
both the observed O$^{+3}$ and Fe$_{\rm gas}$.

Finally, we note that we have observed a wide variety of conditions in the GC.
These would all occur {\it in the same aperture} in an observation of a distant 
starburst galaxy.
One cannot {\it assume} that the highest excitation gas is associated with 
the most massive cluster --- in the GC it most certainly is {\it not}.

\acknowledgments
This work is based on observations made with the {\it Spitzer
Space Telescope}, which is operated by the Jet Propulsion Laboratory,
California Institute of Technology, under a contract with NASA. 
Support for this work was provided by NASA.
JPS acknowledges support from NASA/Ames Research Center Research Interchange Grant NNA05CS33A 
to the SETI Institute.
RHR acknowledges NASA support through LTSA and Spitzer grants.
The IRS was a collaborative venture between Cornell University 
and Ball Aerospace Corporation 
funded by NASA through the Jet Propulsion Laboratory and Ames Research Center.
SMART was developed by the IRS Team at Cornell University and is available 
through the Spitzer Science Center at Caltech.
We thank M. R. Haas and A. G. G. M. Tielens for reading the manuscript 
and we thank the referee for the thoughtful and thorough review
which improved the paper significantly.

{\it Facility:} \facility{Spitzer (IRS)}.

\appendix

\section{Extinction Laws}
The extinction law of Chiar \& Tielens (2006) for the GC
is notable for having a much larger 18 \micron\ silicate feature
(with respect to the 10 \micron\ silicate feature) 
than the older relations of Draine
(Draine \& Lee 1984; Draine 2003a, b).
With the extinction as high as it is in the GC, the choice of 
extinction law for the extinction correction obviously affects 
the various MIR line ratios.

We can test this effect using the three observed H$_2$ lines.
Figure 15 shows flux ratios of the extinction-corrected H$_2$ S(2)/S(1) 
and S(0)/S(2) lines plotted versus the optical depth at 9.6 \micron, $\tau_{9.6}$
(the 10 \micron\ silicate feature peaks at 9.6 \micron\ in 
Chiar \& Tielens' (2006) table for the GC).
Because Draine's (2003a, b) relation is substantially different, we also estimate 
$\tau_{9.6}$ for each position using his law for the diffuse ISM ($R=3.1$) 
and then correct our observed H$_2$ line fluxes for this extinction.
The ratios of the fluxes thus corrected are also plotted in Fig.~15.
We see that the H$_2$ flux ratios corrected with Draine's extinction law 
have a strong dependence on $\tau_{9.6}$ whereas the flux ratios 
corrected with Chiar \& Tielens' GC law have a much smaller dependence on $\tau_{9.6}$.
Since the extinction at 28 \micron\ is due to the tail of the 18 \micron\ silicate feature,
the S(0)/S(1) flux ratio does not show any difference due to extinction law and is not plotted;
the slope is slightly negative, perhaps indicating that the positions with
higher extinction (mostly in or near the Arched Filaments) have warmer H$_2$.

Figures 16 and 17 shows the H$_2$ S(0)/S(1) and S(0)/S(2) flux ratios plotted 
versus the S(2)/S(1) flux ratio 
for our observed fluxes corrected for extinction with the laws from
both Chiar \& Tielens (2006) for the GC and from Draine (2003a, b).
We also plot curves of the theoretical equilibrium ratios for local thermodynamic equilibrium (LTE). 
The observed flux ratios fall far above this line; 
in effect, the S(1) line is much weaker than would be predicted from the S(0)/S(2) ratio.
The LTE curve assumes the equilibrium ortho/para ratio 
(an extra factor of 3 in the statistical weights for the odd levels) --- 
deviations such as seen here have been interpreted as non-equilibrium excitation due to shocks.
However, it is not likely that the observed H$_2$ emission comes from a region of 
constant physical conditions along the line-of-sight to the GC.
The dashed and dotted lines show that the observed ratios can be produced by simple two-component models:
a cooler molecular cloud that produces most of the H$_2$ S(0) flux and
a warmer PDR that produces most of the H$_2$ S(2) flux. 
However, we see that the flux ratios corrected with the Draine (2003a, b) extinction law 
require a much larger high-temperature component than 
the flux ratios corrected with the Chiar \& Tielens (2006) extinction law.
We conclude from Figs. 15 to 17 that the Chiar \& Tielens (2006) GC extinction law 
gives a better fit to the H$_2$ line flux ratios than does the Draine (2003a, b) extinction law.

\clearpage

\input tab1.tex

\input tab2.tex

\input tab3.tex

\input tab4.tex

\clearpage

\begin{figure}
\epsscale{1.0}
\plottwo{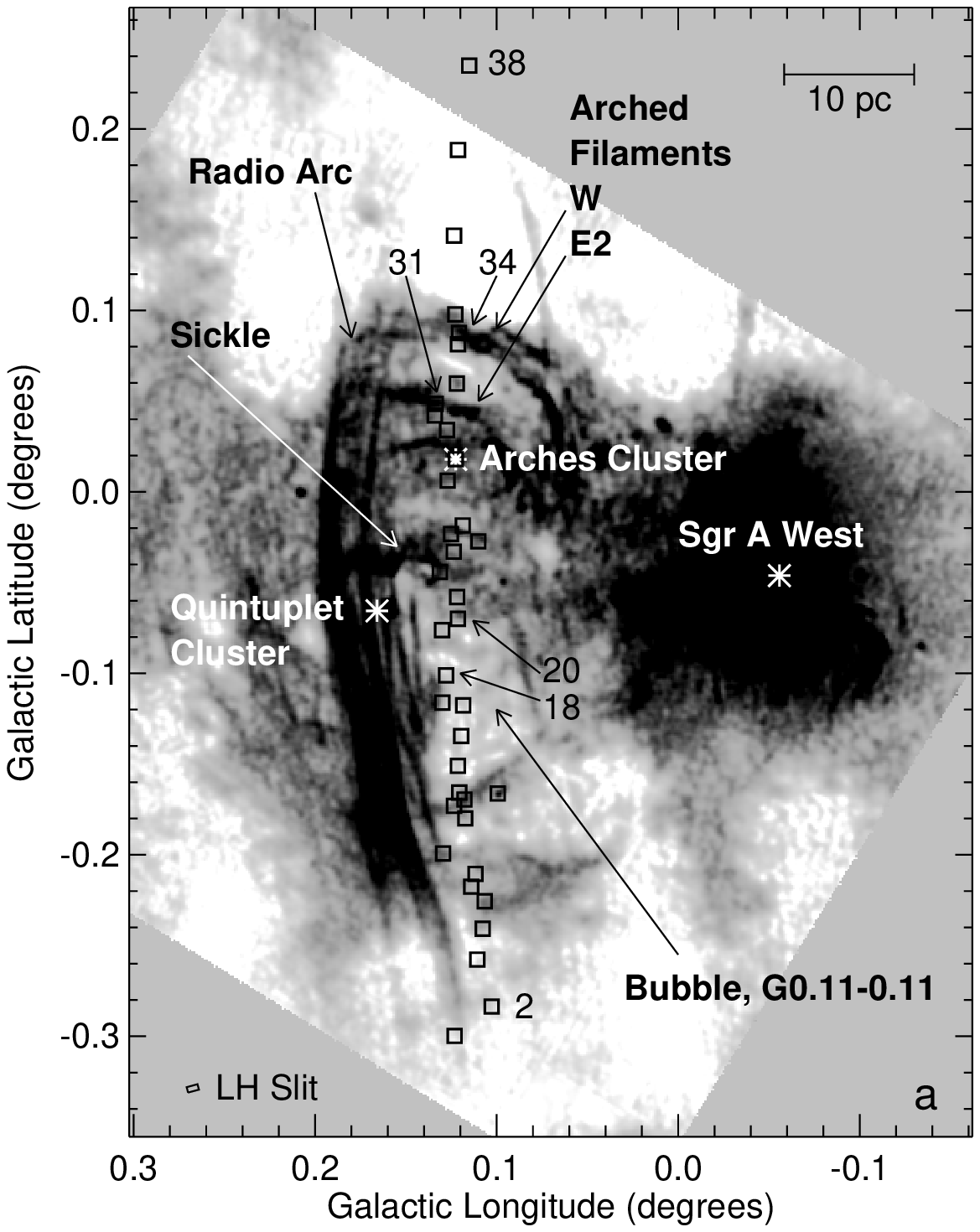}{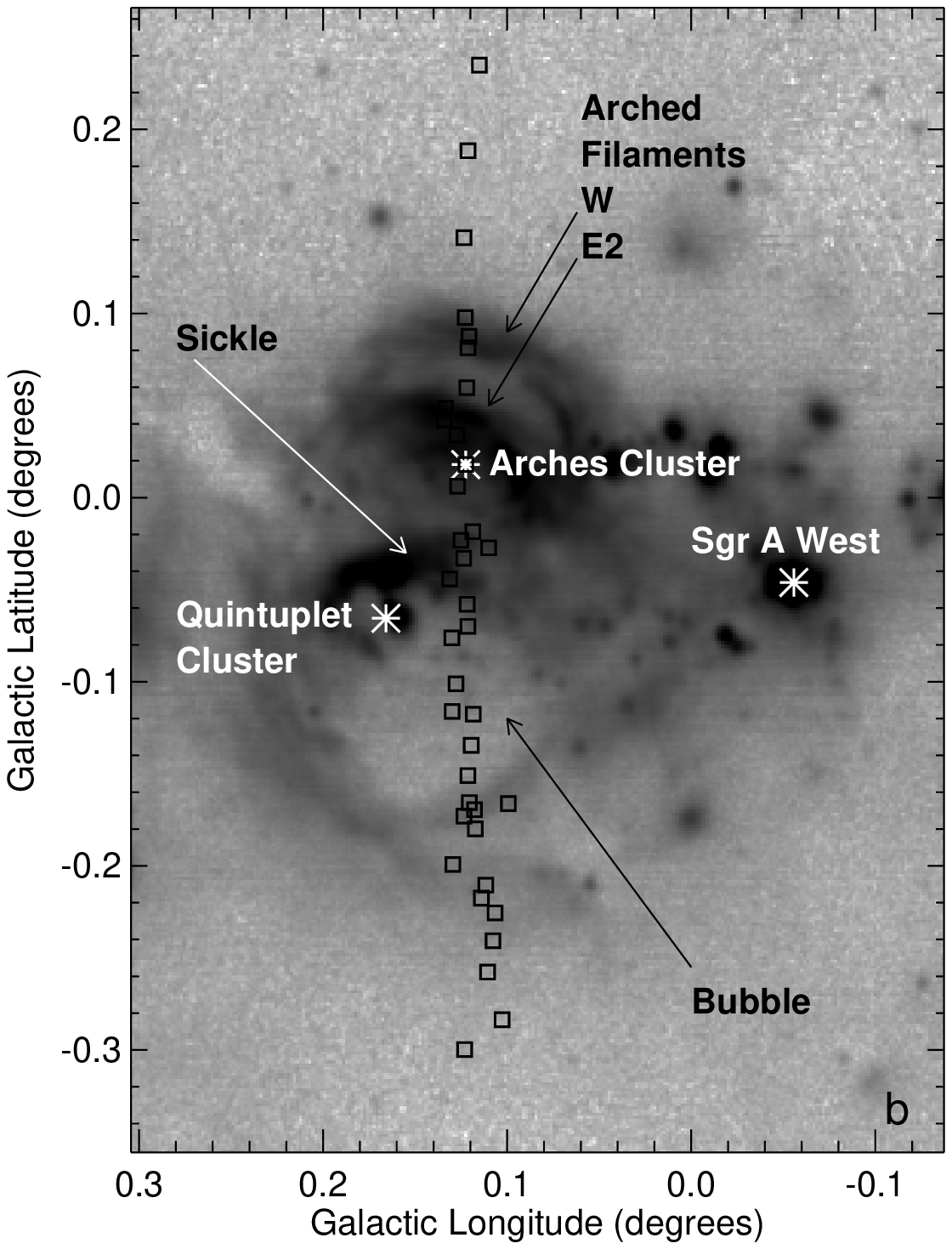}
\caption{Images of the GC region with the observed positions indicated by square boxes.
Other positions of interest are labeled.
(a) Radio continuum (log log scale) imaged at 21 cm 
by Yusef-Zadeh \& Morris (1987a) with $11''$ resolution.
The Radio Arc and part of the Sgr A region are nonthermal; the rest of the radio emission is 
thermal bremsstrahlung. 
A few positions are labeled with numbers (see Table~1).
Also indicated is the size ($11''$ by $22''$) and orientation of the LH slit.
Including the small maps, the actual size of the observed positions is $11''$ by $30''$ for LH and 
$13''$ by $14''$ for SH. 
(b) MSX Band E (21 \micron) image (log scale) from Price et al. (2001). 
\label{fig1}
}
\end{figure}

\begin{figure}
\epsscale{1.0}
\plotone{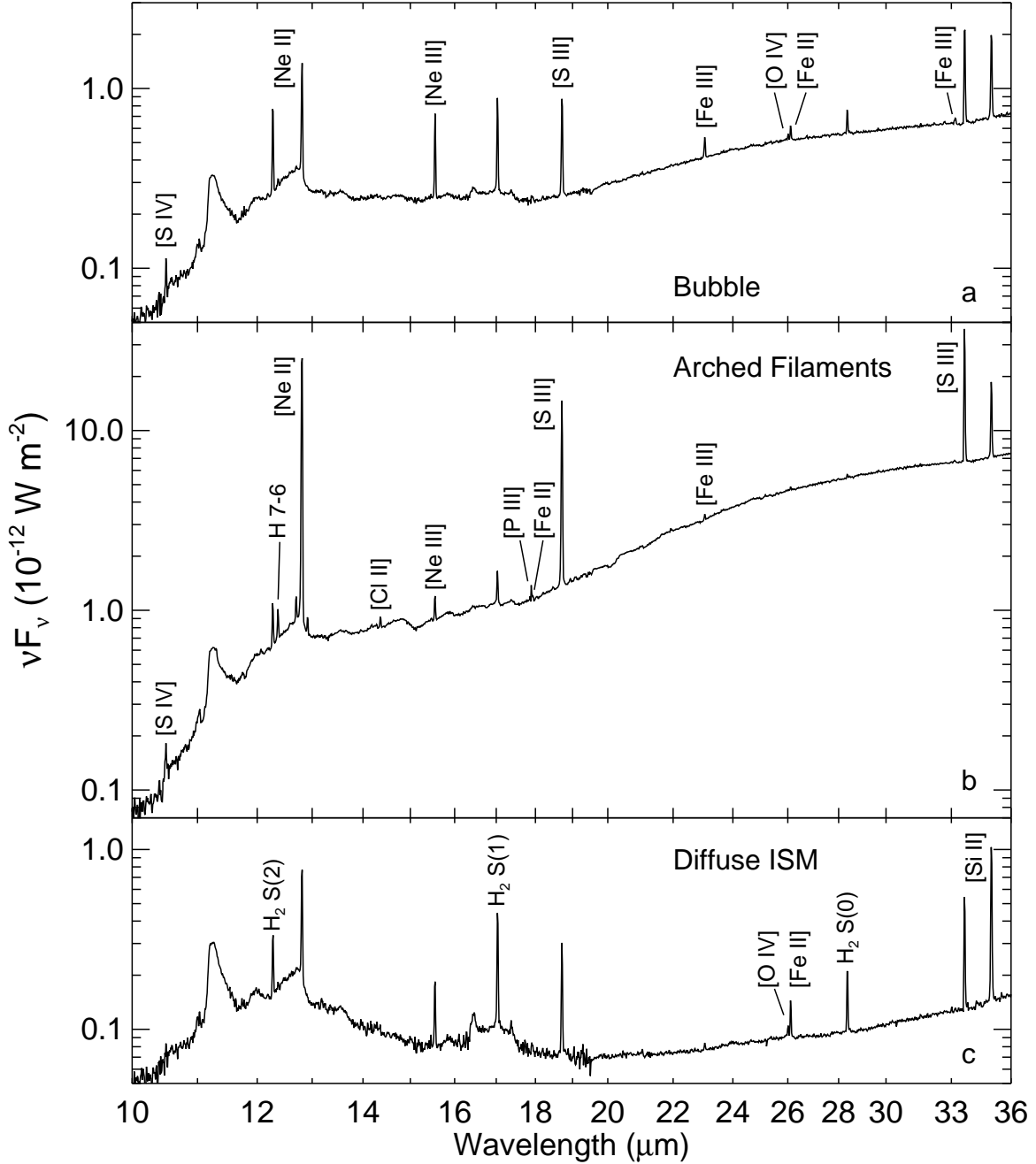}
\caption{Three typical spectra.
Lines representative of each type of location are marked
(almost all lines are detected at each position).
The broad features at 11.0, 11.2, 12.0, 12.6, 13.5, 14.2, and 16.3 -- 17.5 \micron\ 
are due to polycyclic aromatic hydrocarbons (PAHs).
(a) The center of the Bubble (Position 18).
(b) GC \ion{H}{2} regions (Position 34 in the Arched Filaments).
The spikes on either side of the [\ion{Ne}{2}] 12.8 \micron\ line are ghosts.
(c) The diffuse ISM at the extreme distance from the Galactic Plane 
(Position 38).
}
\end{figure}

\begin{figure}
\epsscale{1.0}
\plotone{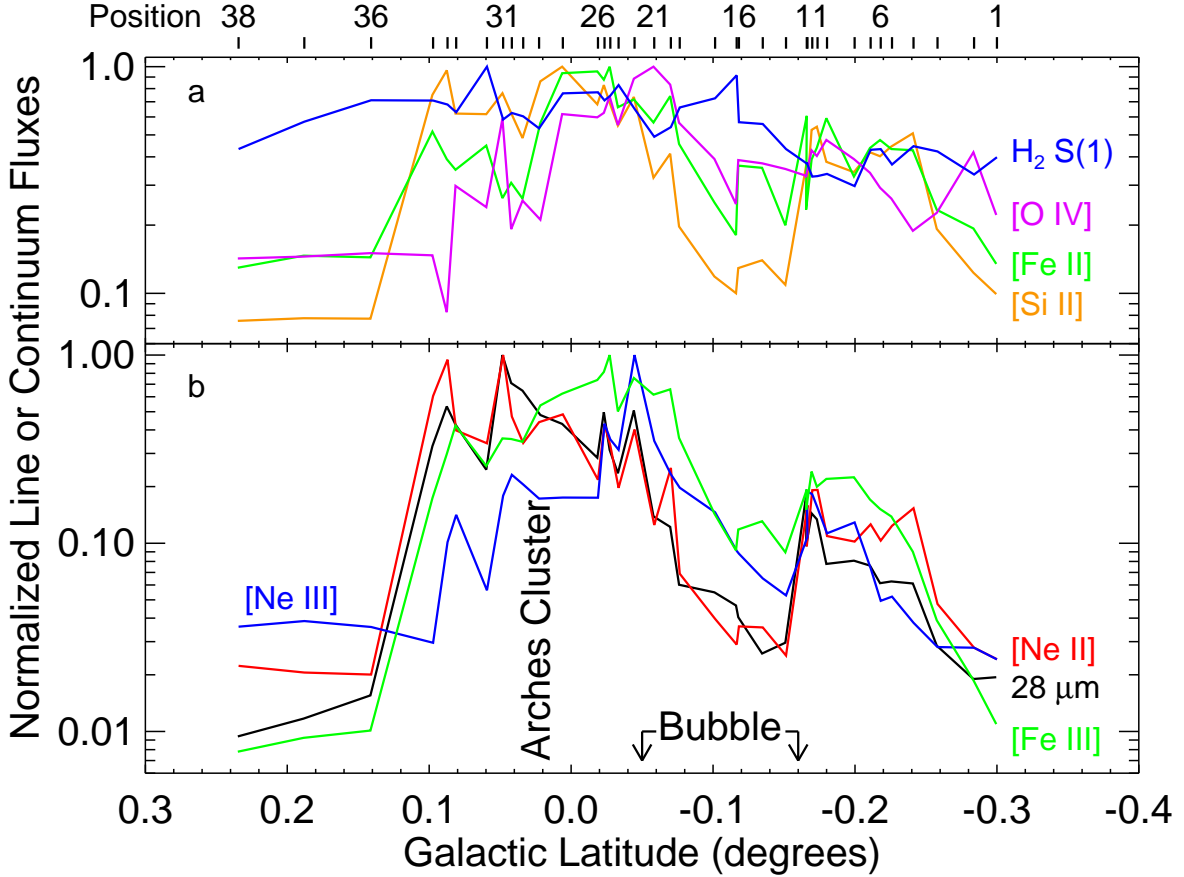}
\caption{Normalized fluxes for all positions as a function of Galactic latitude.
(a) The H$_2$ S(1) 17.0 \micron\ line, the \ion{H}{2} region and PDR lines of [\ion{Si}{2}] 
and [\ion{Fe}{2}] and the high excitation [\ion{O}{4}] line are plotted.
The normalizing fluxes (SH fluxes scaled to the LH beam =  248 square arcsec) are 
H$_2$ S(1): $1.45 \times 10^{-15}$ W m$^{-2}$,
[\ion{Si}{2}] 34.8 \micron: $ 2.33 \times 10^{-14}$ W m$^{-2}$,
[\ion{Fe}{2}] 26.0 \micron: $ 6.93 \times 10^{-16}$ W m$^{-2}$,
and [\ion{O}{4}] 25.9 \micron: $ 1.49 \times 10^{-16}$ W m$^{-2}$.
(b) \ion{H}{2} region lines of [\ion{Ne}{2}], [\ion{Ne}{3}], and [\ion{Fe}{3}] are plotted, along with 
the 28 \micron\ continuum from the PDR-heated molecular clouds.
The normalizing fluxes are 
28 \micron\ continuum: 97 Jy,
[\ion{Ne}{2}] 12.8 \micron: $5.09  \times 10^{-14}$ W m$^{-2}$,
[\ion{Ne}{3}] 15.6 \micron: $5.65  \times 10^{-15}$ W m$^{-2}$,
and [\ion{Fe}{3}] 22.9 \micron: $1.64 \times 10^{-15}$ W m$^{-2}$.
The [\ion{S}{3}] lines are not plotted since they have very similar distributions 
to the [\ion{Ne}{2}] line.
}
\end{figure}

\begin{figure}
\epsscale{1.0} 
\plotone{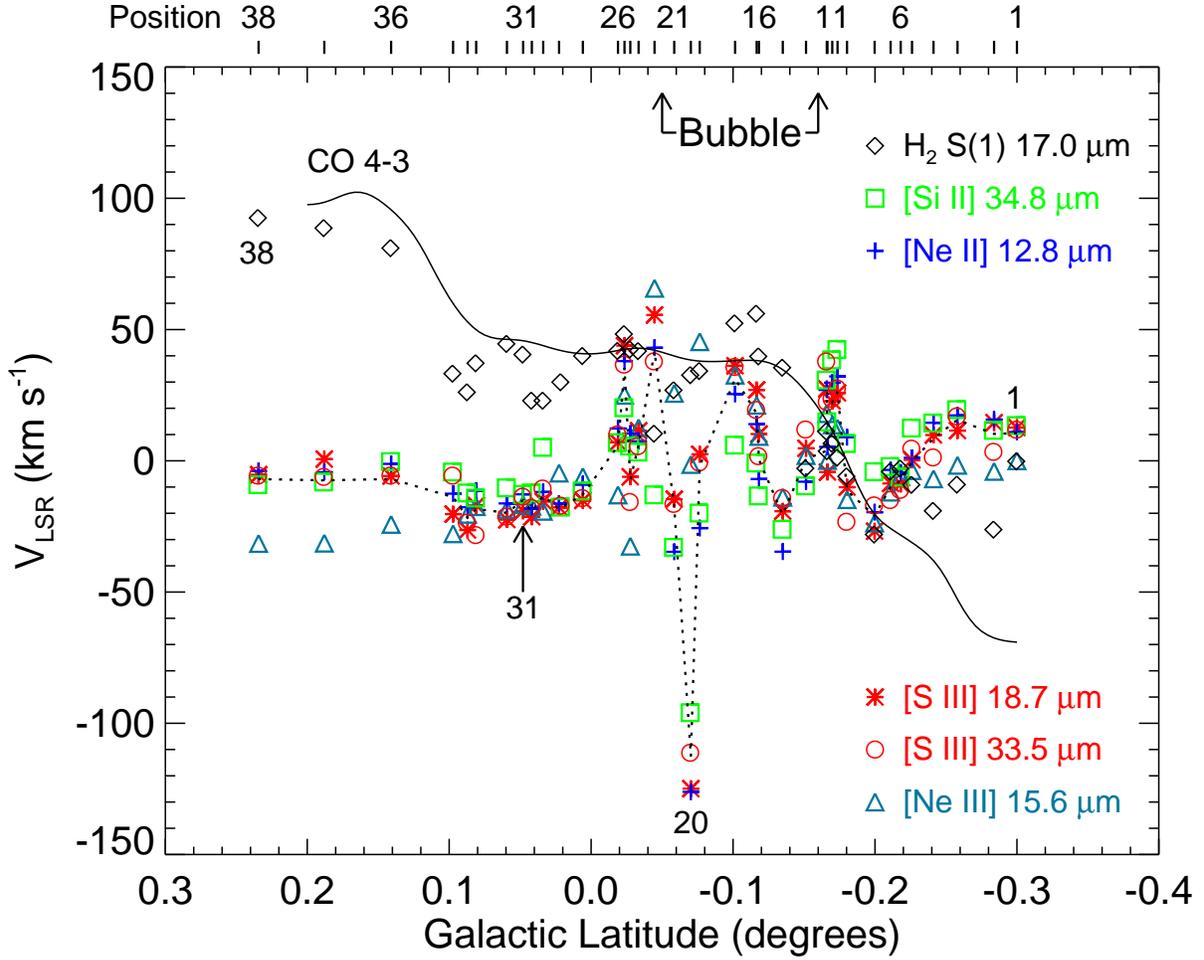}
\caption{Radial velocities $V_{\rm LSR}$ as a function of Galactic latitude.
A few positions are labeled with numbers.
The velocities of the ionized lines are normalized to $-15$ km s$^{-1}$ 
at Position 31 in the Arched Filaments,
based on radio recombination line measurements there (Lang et al. 2001).
The dotted line is the average of the [\ion{Si}{2}], [\ion{Ne}{2}], and [\ion{S}{3}] velocities 
at each position (Table~1).
The average H$_2$ S(1) line velocity is normalized to the average of the CO J = 4 -- 3 velocities 
measured by Martin et al. (2004)
and the solid line is these velocities at each position.
For most positions, the measured uncertainties for the line centers  
are $\lesssim 2$  km s$^{-1}$ for the brightest lines: [\ion{Ne}{2}], [\ion{S}{3}], and [\ion{Si}{2}],
and $<10$  km s$^{-1}$ for [\ion{Ne}{3}] and H$_2$ S(1).
The dispersions in the difference of the measured velocities for pairs of bright lines 
are $\sim 10 - 20$  km s$^{-1}$;
this probably represents the actual measurement uncertainty for the brightest lines.
}
\end{figure}

\begin{figure}
\epsscale{1.0}
\plotone{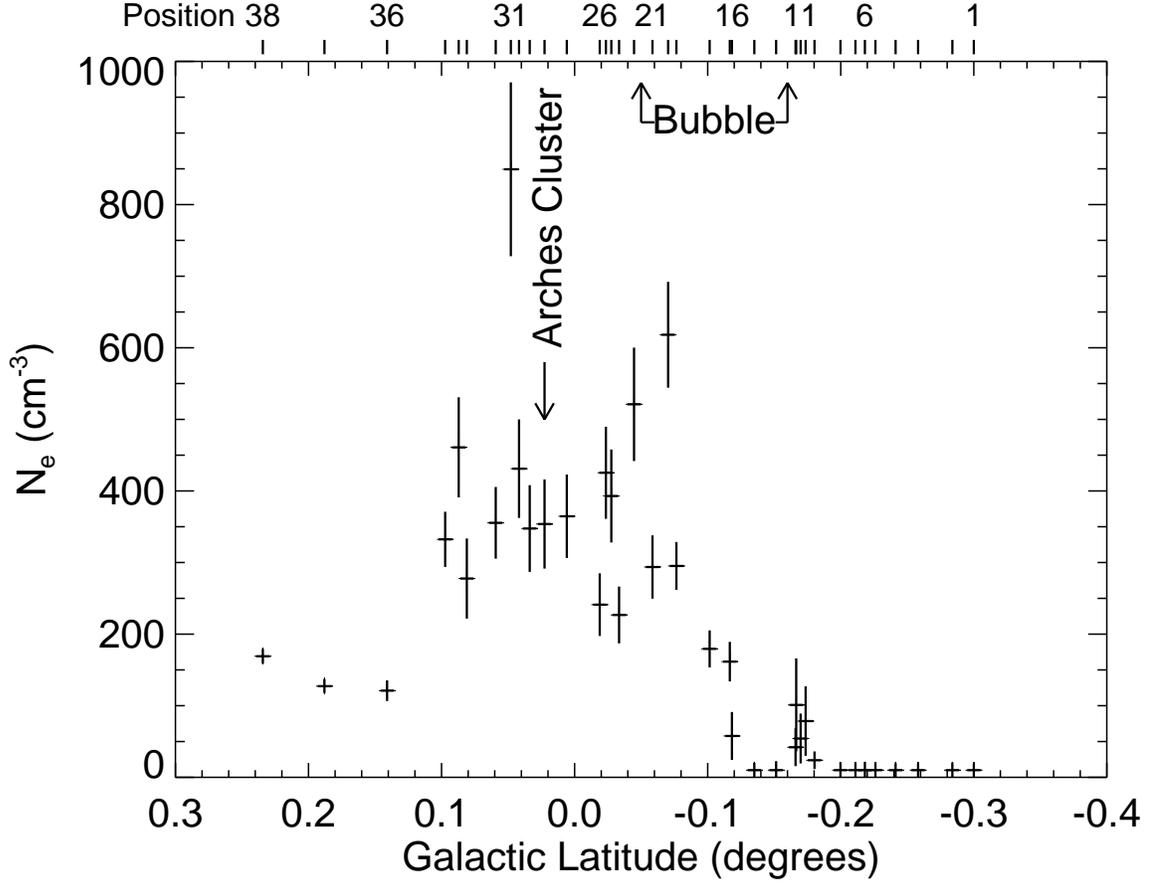}
\caption{Electron density ($N_e$) estimated from the  
[\ion{S}{3}] 18.7/33.5 \micron\ line flux ratio plotted as a function of Galactic latitude.
The Galactic latitudes of the Arches Cluster and the Bubble are marked.
Most of the error bars are due to uncertainties in the extinction. 
For 8 of the 10 positions with $N_e = 10$ cm$^{-3}$ (Table~1), 
the [\ion{S}{3}] 18.7/33.5 \micron\ line flux ratio computed with the original estimated extinction 
was lower than the theoretical lower limit for $T_e = 6000$~K;
for these positions the extinction was increased until a density of 10 cm$^{-3}$ could be obtained.}
\end{figure}

\begin{figure}
\epsscale{1.0}
\plotone{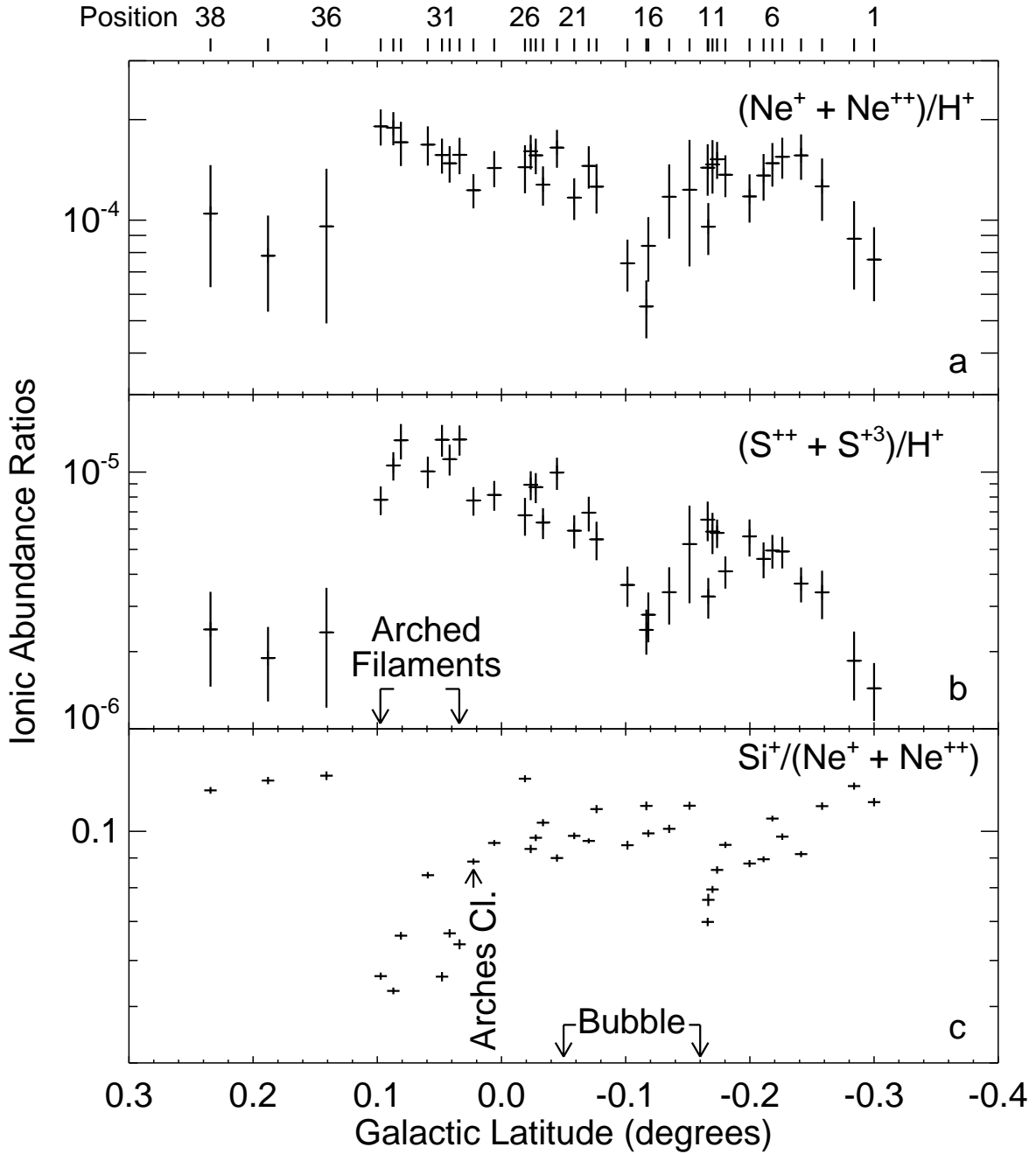}
\caption{Ionic abundances ratios plotted as a function of Galactic latitude.
(a) The (Ne$^{+}$ + Ne$^{++}$)/H$^{+}$ ratio. 
(b) The (S$^{++}$ + S$^{+3}$)/H$^{+}$ ratio. 
(c) The Si$^{+}$/(Ne$^{+}$ + Ne$^{++}$) ratio. 
}
\end{figure}

\begin{figure}
\epsscale{1.0}
\plotone{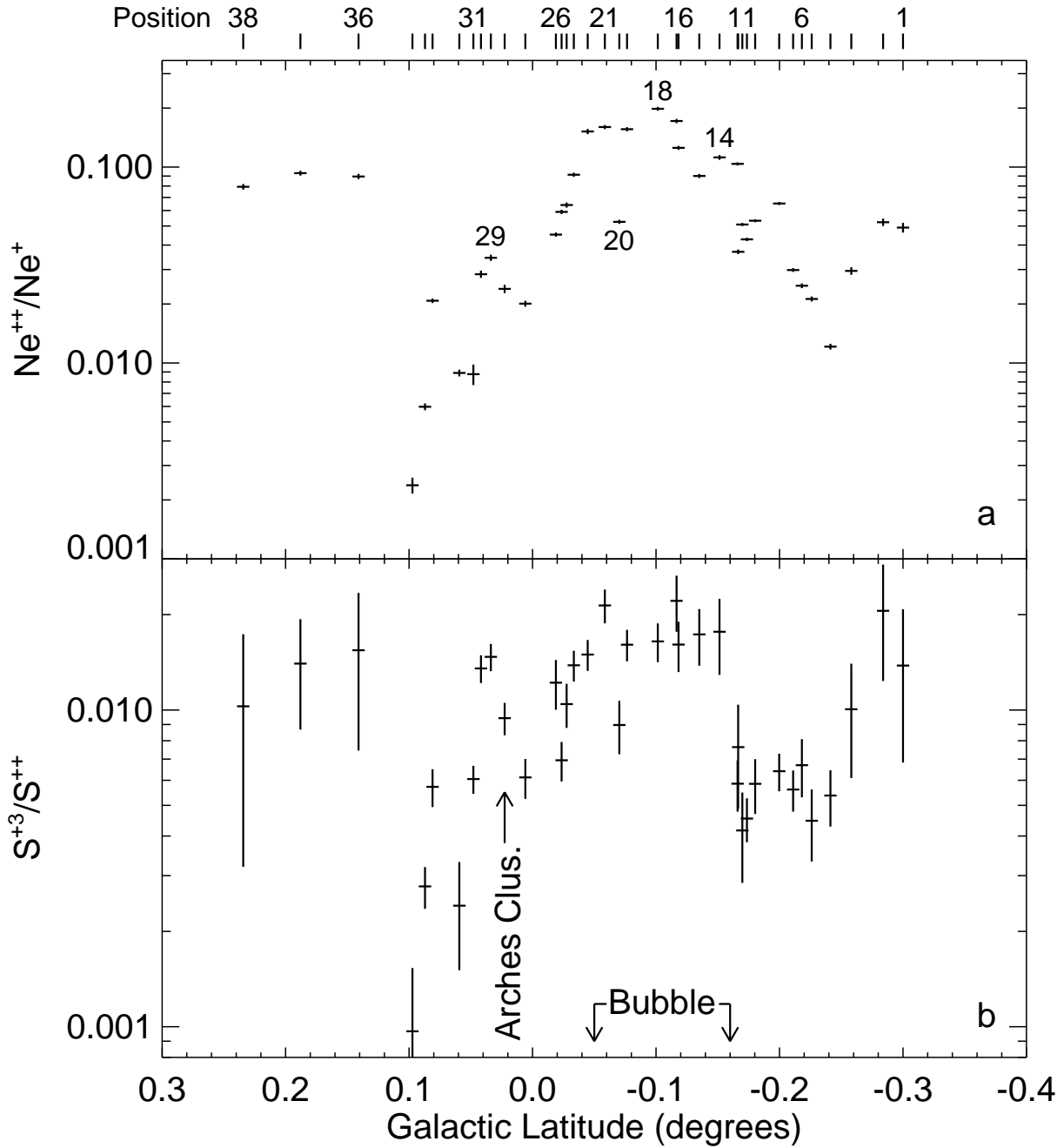}
\caption{Ratios indicative of gas excitation plotted as a function of Galactic latitude.
(a) The Ne$^{++}$/Ne$^{+}$ ratio. 
(b) The S$^{+3}$/S$^{++}$ ratio. 
This ratio has large error bars because the [\ion{S}{4}] line suffers so much extinction.
}
\end{figure}

\begin{figure}
\epsscale{1.0}
\plotone{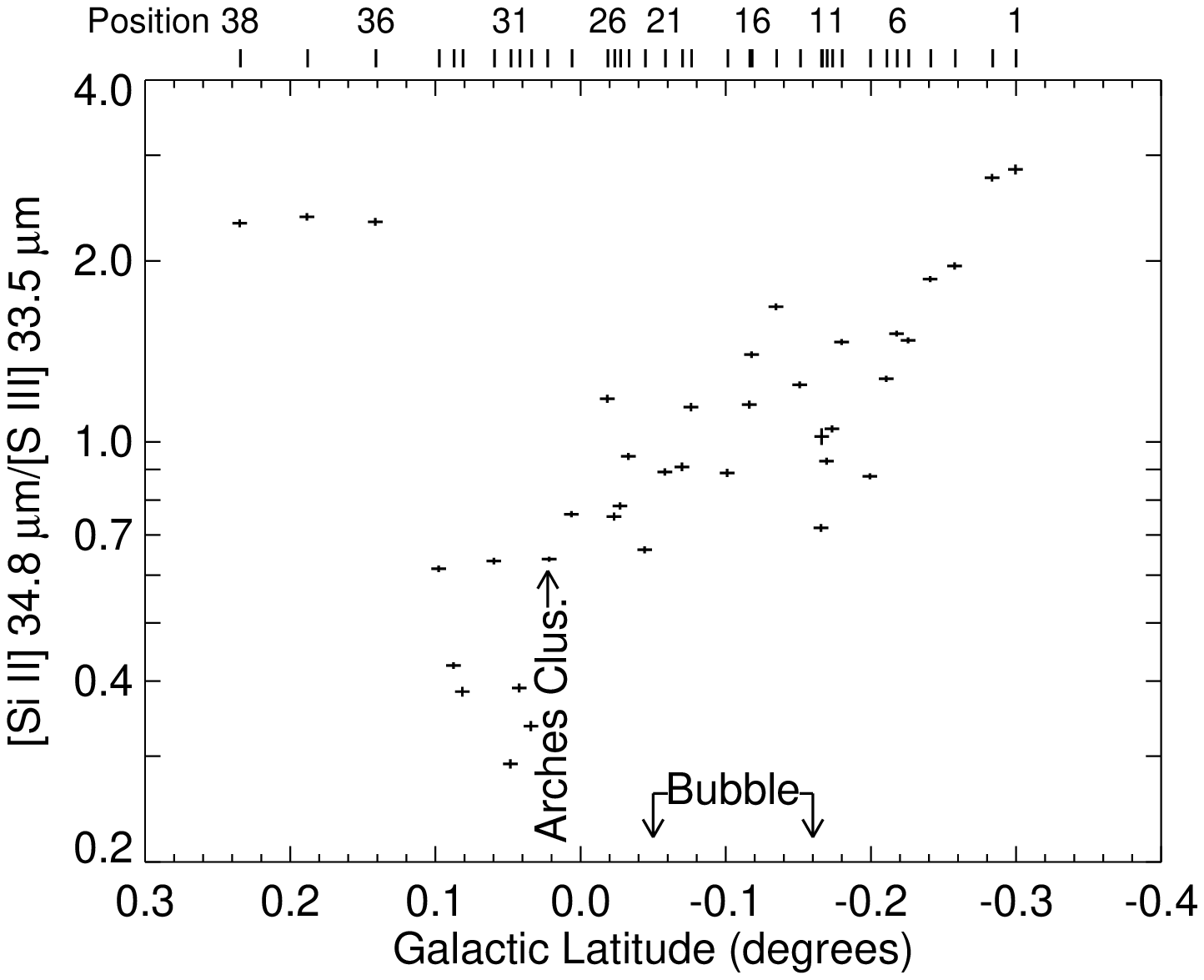}
\caption{The [\ion{Si}{2}] 34.8 \micron/[\ion{S}{3}] 33.5 \micron\ line flux ratio 
plotted as a function of Galactic latitude.
}
\end{figure}

\begin{figure}
\epsscale{1.0}
\plotone{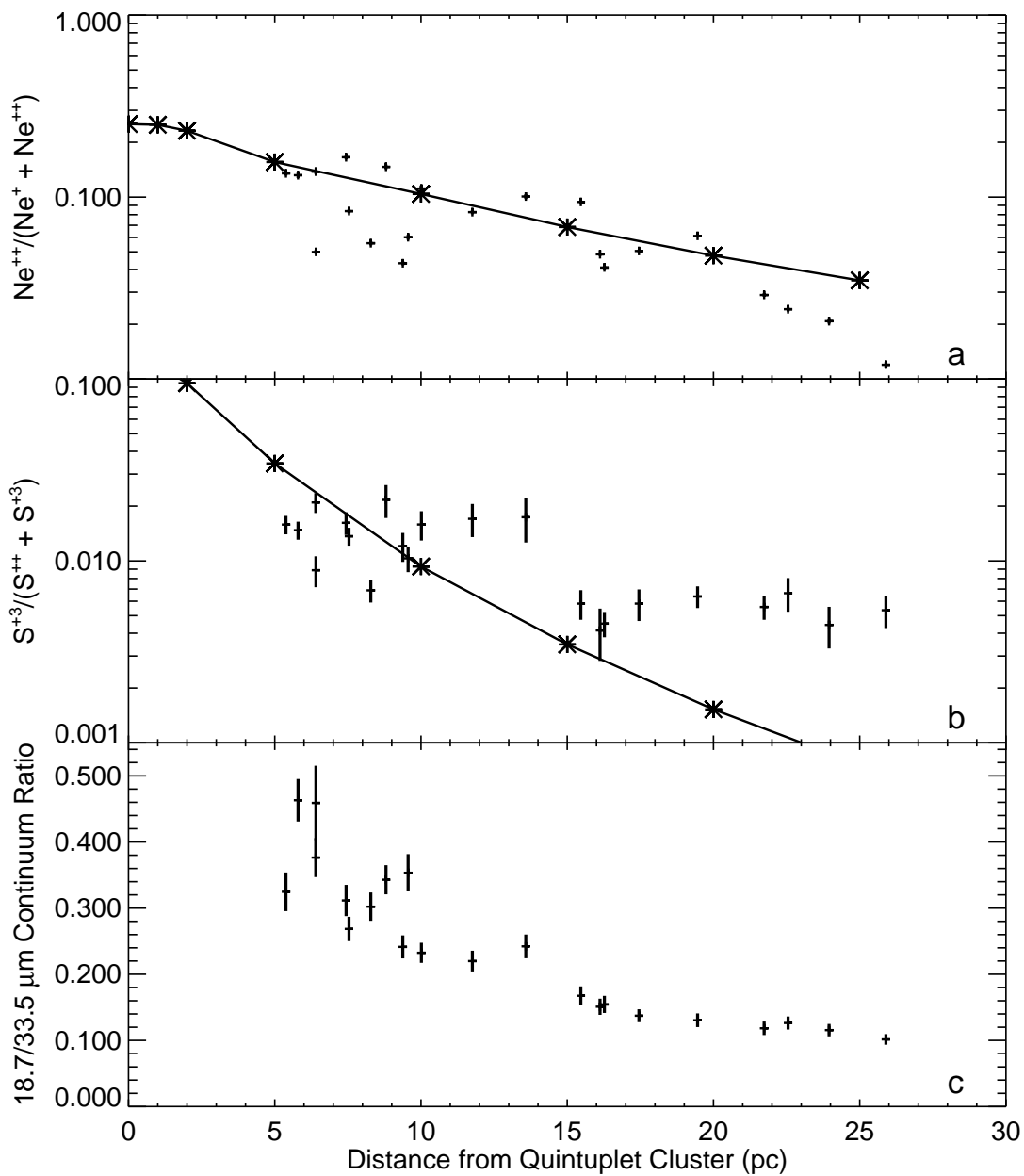}
\caption{Ionization fractions and continuum flux ratios 
affected by dilution of the radiation field, 
plotted as a function of distance from the Quintuplet Cluster on the plane of the sky.
The measured ratios are for Positions 4 -- 26, excluding Position 13,
which may be a separate compact \ion{H}{2} region.
In addition, 
ratios computed from models calculated using CLOUDY (Ferland et al. 1998)
are plotted as asterisks connected by solid lines (see text).
(a) Ne$^{++}$/(Ne$^+$ + Ne$^{++}$).
(b) S$^{+3}$/(S$^{++}$ + S$^{+3}$).
(c) Ratio of the extinction-corrected continuum fluxes at 18.7 and 33.5 \micron.
}
\end{figure}

\begin{figure}
\epsscale{1.0}
\plotone{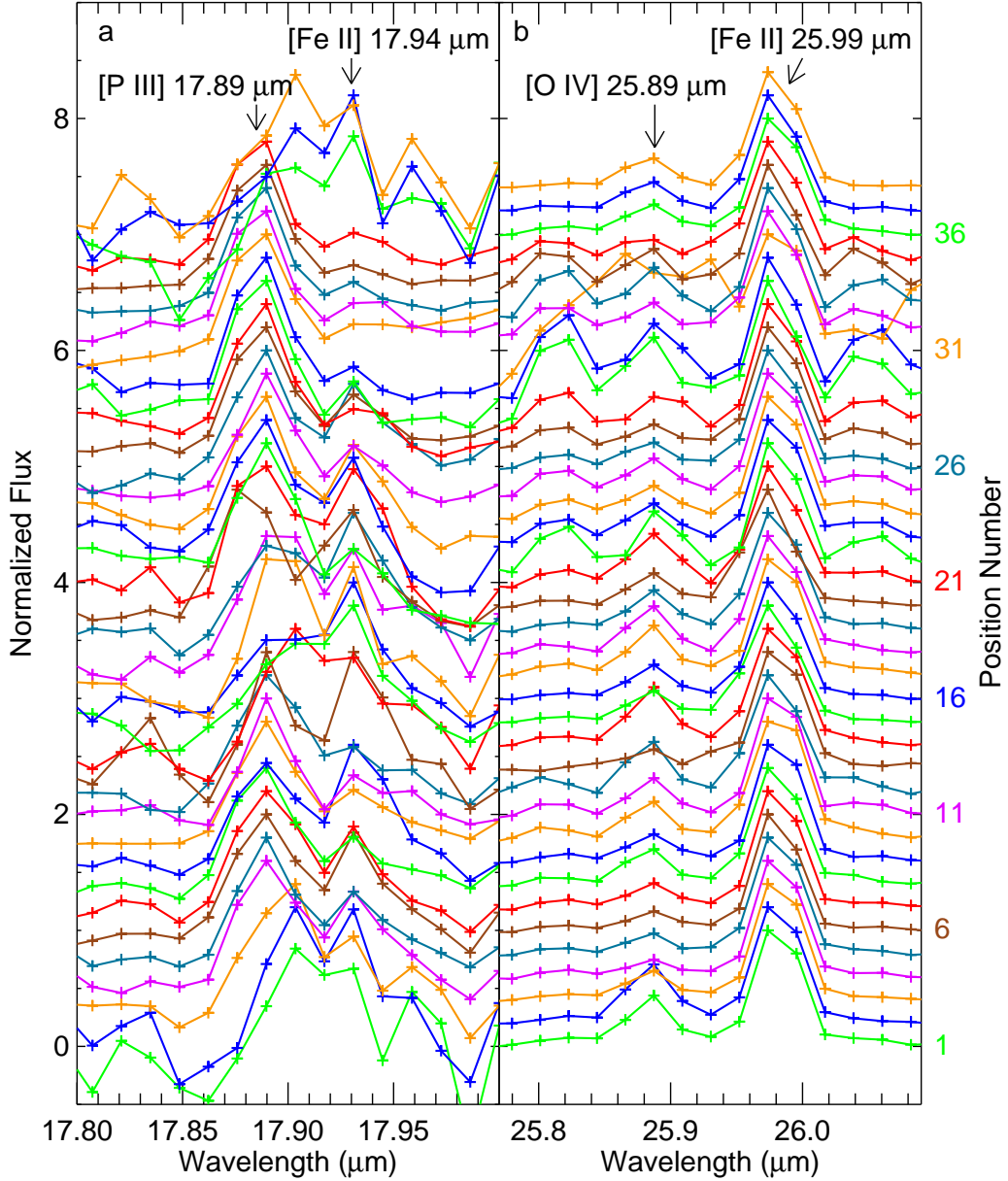}
\caption{Normalized spectrum segments (continuum subtracted) showing weak 
but important lines (see Fig.~2 for typical strong lines). 
The spectra are plotted in order of position number (Table~1) from bottom to top
and are labeled every 5 positions.
(a) The 17.8 -- 18.0 \micron\ spectral region has the  
[\ion{P}{3}] 17.89 \micron\ and [\ion{Fe}{2}] 17.94 \micron\ \ion{H}{2} region lines.
(b) The 25.77 -- 26.09 \micron\ spectral region has the 
high excitation [\ion{O}{4}] 25.89 \micron\ line and the 
[\ion{Fe}{2}] 25.99 \micron\ PDR line.
}
\end{figure}

\begin{figure}
\epsscale{1.0}
\plotone{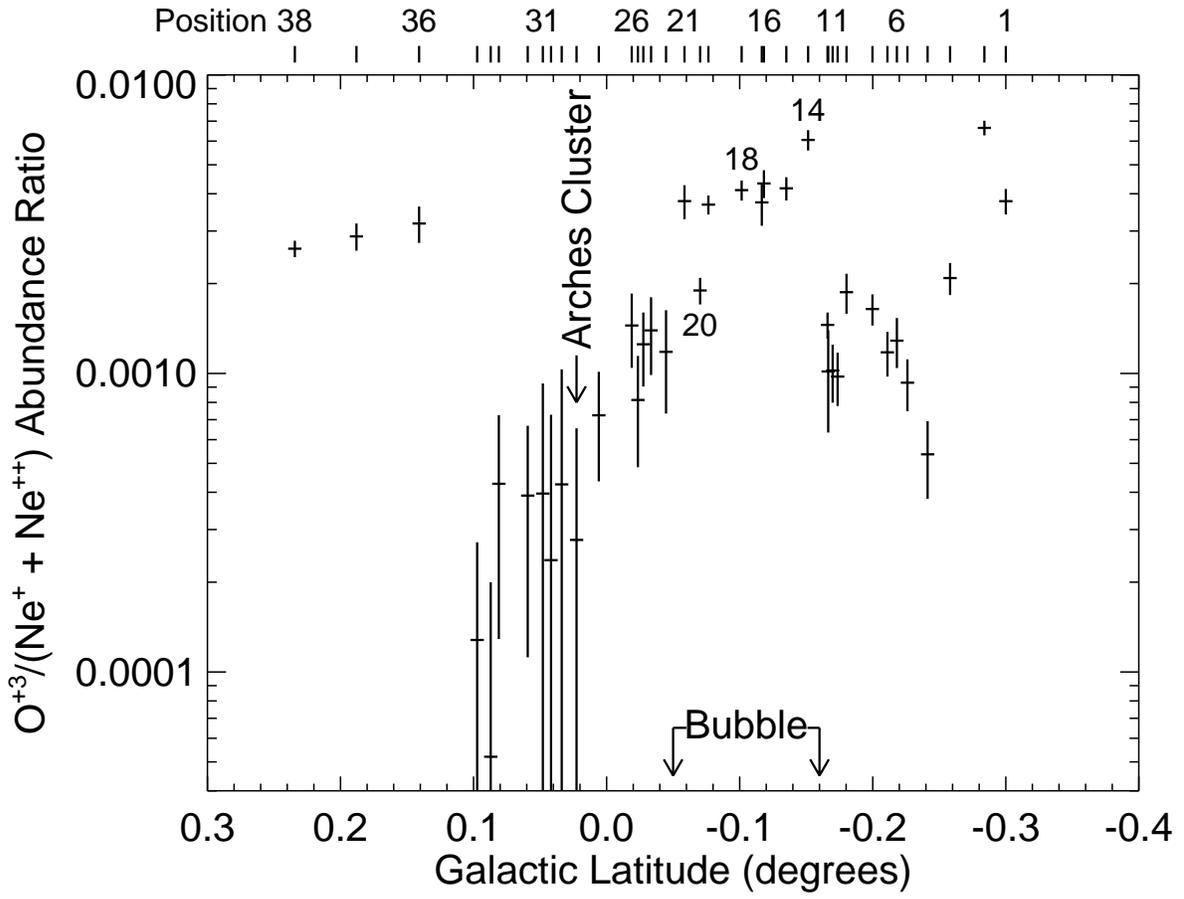}
\caption{The O$^{+3}$/(Ne$^{+}$ + Ne$^{++}$) ratio 
plotted as a function of Galactic latitude.
}
\end{figure}

\begin{figure}
\epsscale{0.8}
\plotone{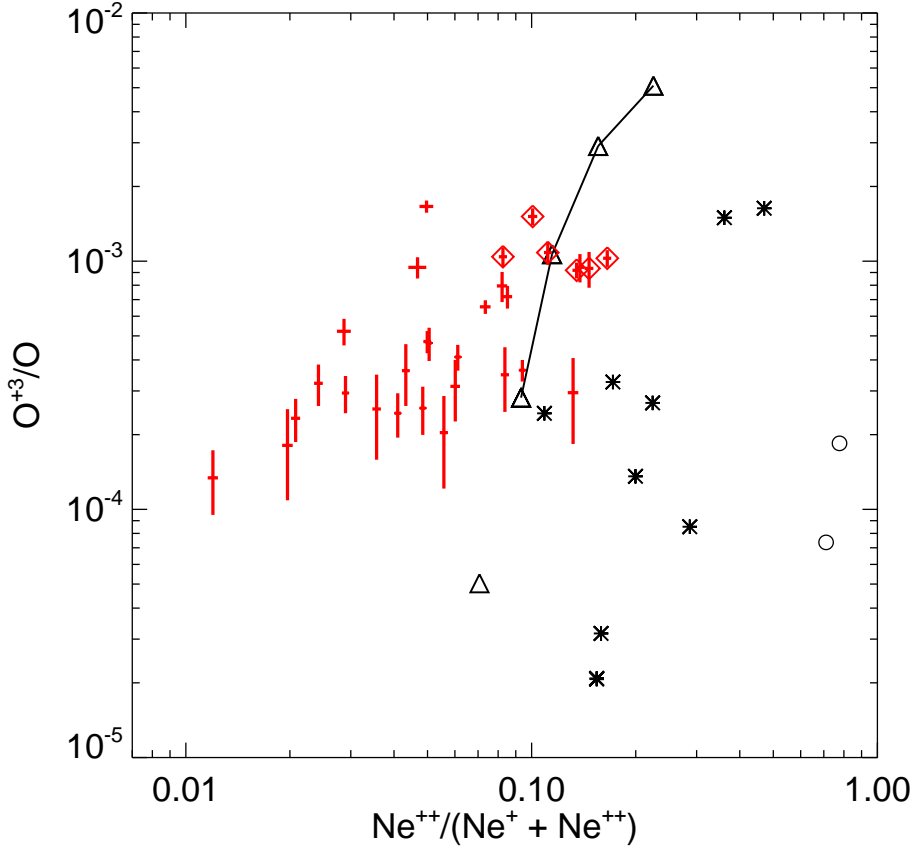}
\caption{The O$^{+3}$/O ratio plotted versus the Ne$^{++}$/(Ne$^{+}$ + Ne$^{++}$) ratio 
from various \ion{H}{2} region models (see text).
An estimate of the observed O$^{+3}$/O ratio is also plotted as red crosses with error bars, 
where the estimate is calculated by multiplying the measured  
O$^{+3}$/(Ne$^{+}$ + Ne$^{++}$) ratio by the Ne/O$ = 0.25$ ratio found  
for the Orion Nebula (Simpson et al. 2004).
The six positions with diamonds overplotted are Positions 14 -- 19 in the Bubble.
The open circles are the CLOUDY models with the Sternberg et al. (2003) atmospheres 
for $T_{\rm eff} = 50$ kK
(no model with a cooler ionizing SED produced enough O$^{+3}$ to appear on this plot.
The asterisks are CLOUDY models ionized by multiple component black bodies; 
the values of $T_{\rm eff}$ used were 30, 35, and 40 kK for the cool components and 
$10^5$ K (white dwarfs) or $10^6$ K (X-rays) for the hot components.
The triangles are models with ionizing SEDs from Sternberg et al. (2003) ($T_{\rm eff} = 36$ kK)
for photon energies $E < 4.013$ ryd and increased by arbitrary amounts for $E > 4.013$ ryd
(see text).
The models represented by the four triangles connected by lines all have SEDs 
with emission in the He$^{++}$ continuum ($E > 4.013$ ryd)
of a factor of $10^5$ larger than the continuum at 4.0 ryd.
The four models have $R_i = 2$, 5, 10, and 15 pc (top to bottom).
Models with completely unphysical SEDs can produce ratios approaching 
the observed ratios, but these models do not have the decrease in excitation  
with distance from the ionizing stars as is observed.
}
\end{figure}

\begin{figure}
\epsscale{1.0}
\plotone{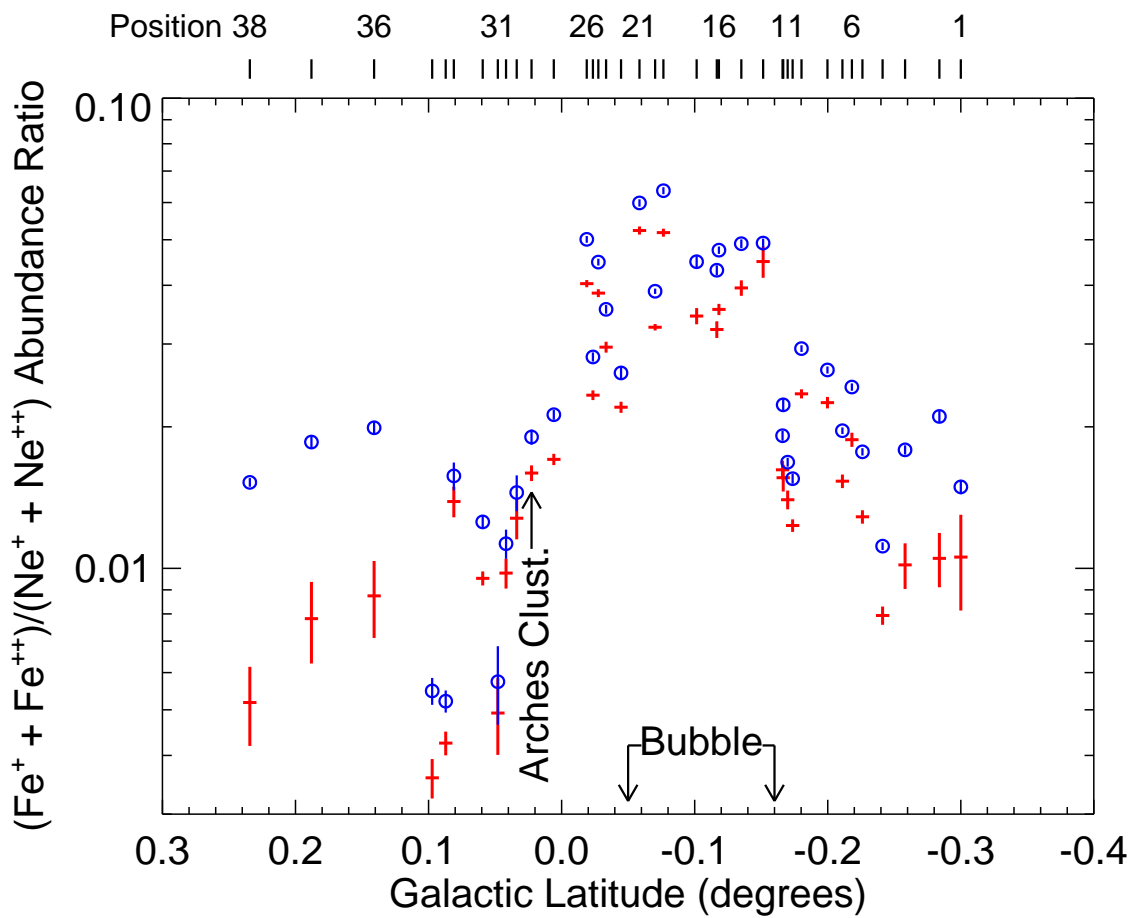}
\caption{The (Fe$^{+}$~+~Fe$^{++}$)/(Ne$^{+}$~+~Ne$^{++}$) ratio 
plotted as a function of Galactic latitude.
The red crosses with larger error bars use the [\ion{Fe}{2}] 17.94 \micron\ line 
to measure the Fe$^+$ abundance and the blue 
open circles with the tiny error bars use the [\ion{Fe}{2}] 25.99 \micron\ line.
The 25.99 \micron\ line is additionally formed in PDRs at much lower $T_e$ than 
the 6000~K used here, 
whereas the 17.94 \micron\ line is exclusively formed in \ion{H}{2} regions 
like the [\ion{Fe}{3}], [\ion{Ne}{2}], and [\ion{Ne}{3}] lines.
Thus, the crosses are more likely to represent the abundance ratios 
of the \ion{H}{2} region gas.
}
\end{figure}

\begin{figure}
\epsscale{1.0}
\plotone{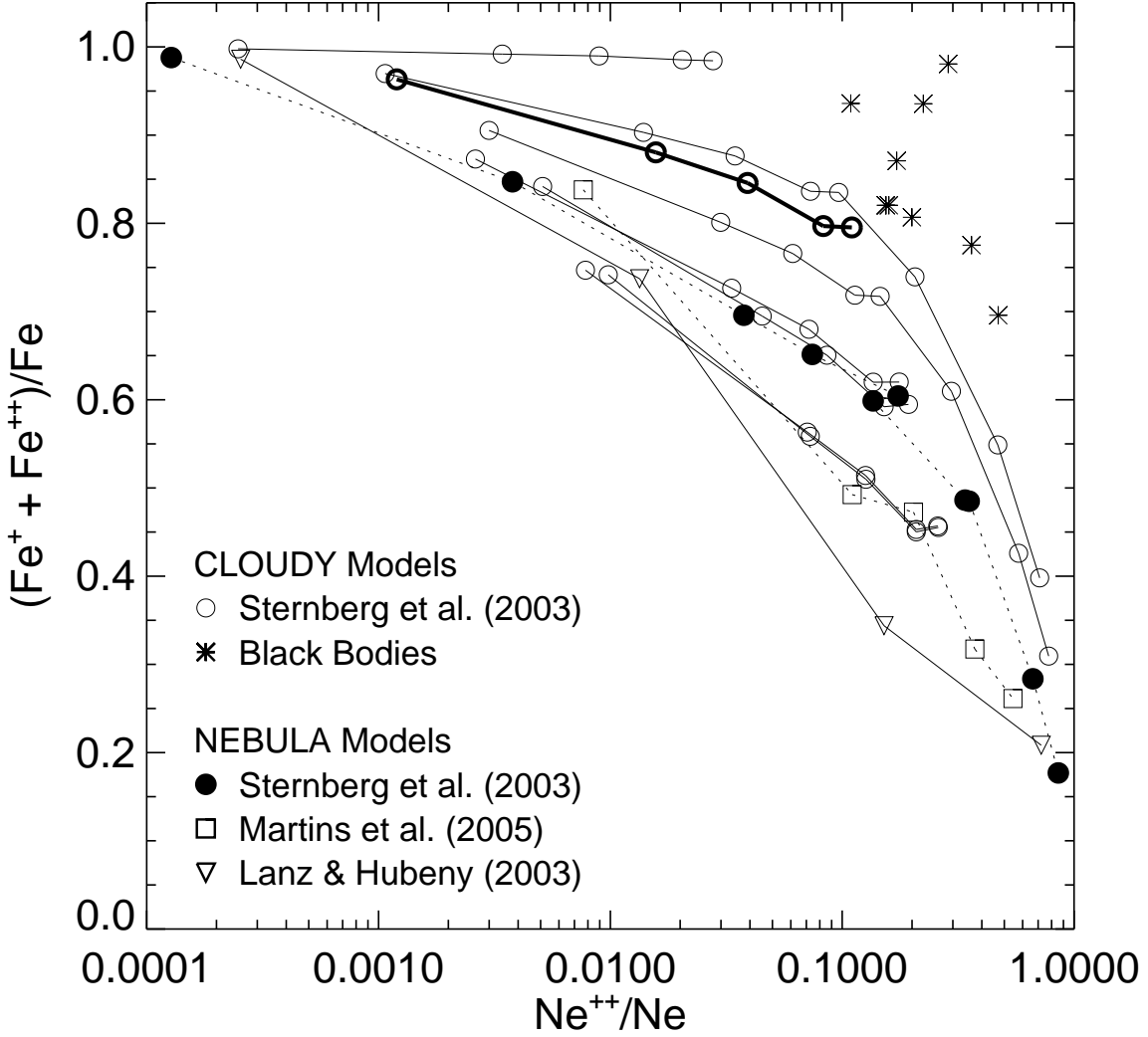}
\caption{The (Fe$^+$ + Fe$^{++}$)/Fe ratio plotted versus the Ne$^{++}$/Ne ratio 
for various \ion{H}{2} region models (see text).
Such models can be used to estimate the correction for higher ionization stages of Fe
in a determination of the abundance of gas-phase iron.
The NEBULA models are for \ion{H}{2} regions with $N_{\rm Lyc} = 10^{49}$ s$^{-1}$ and 
$R_i = 0$,
whereas the CLOUDY models have $N_{\rm Lyc}$ ranging from $10^{49}$ to $10^{51}$ s$^{-1}$ 
and $R_i$ ranging from 0.01 to 10 pc.
The heavy solid line and open circles are the CLOUDY models with Sternberg et al. (2003) 
supergiant SEDs, $T_{\rm eff} = 32 - 37$ kK, $N_{\rm Lyc} = 10^{51}$ s$^{-1}$, 
and $R_i = 10$ pc (Log $U = -2.05$). 
These are the models most likely to represent the regions of the Bubble ionized 
by the Quintuplet Cluster and the regions of the Arched Filaments ionized by 
the Arches Cluster.
}
\end{figure}

\begin{figure}
\epsscale{1.0}
\plotone{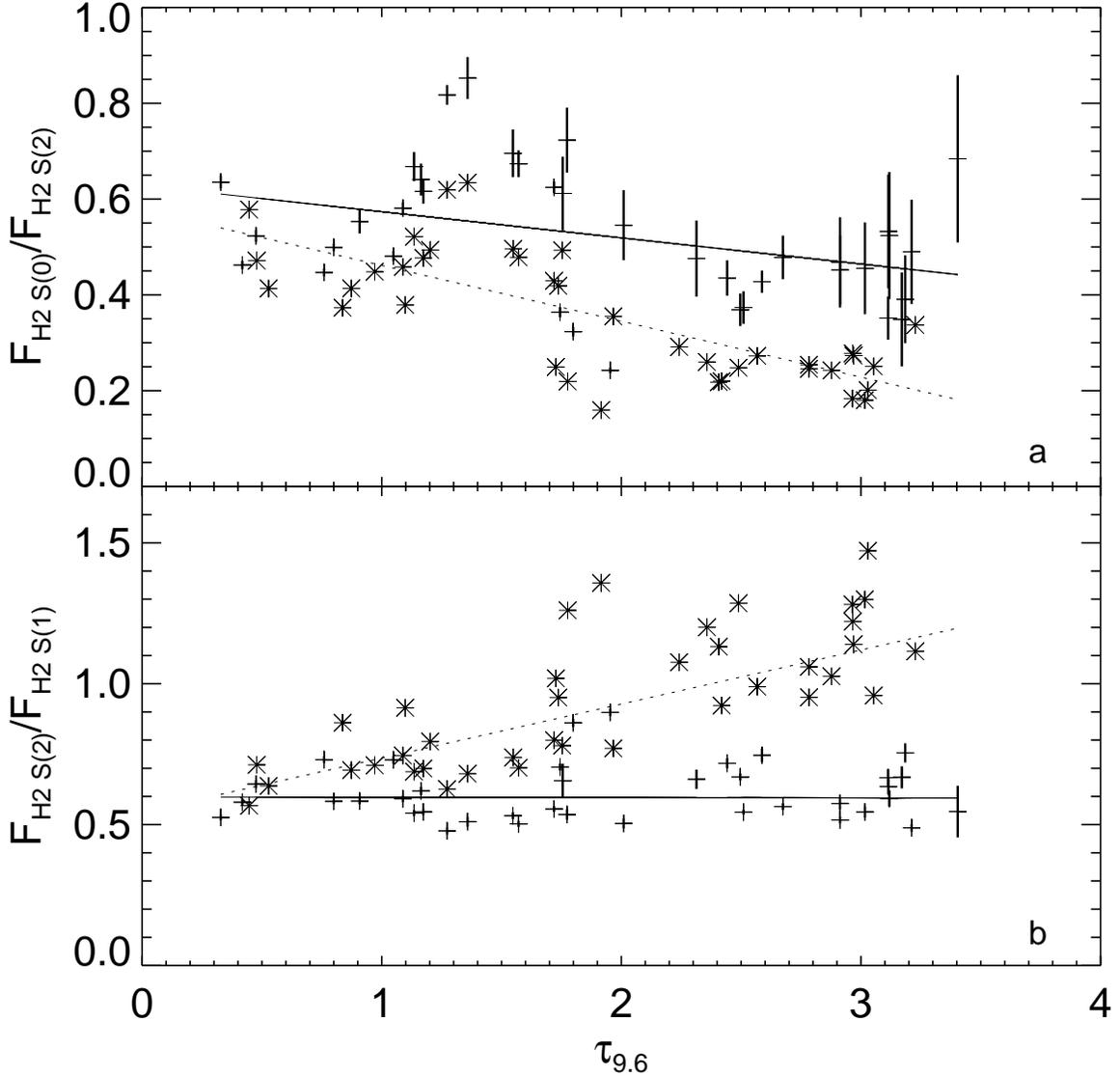}
\caption{Molecular hydrogen ratios plotted versus optical depth 
at 9.6 \micron, $\tau_{9.6}$,
where $\tau_{9.6}$ is determined by fitting the ratio of continuum fluxes at 10 and 14 \micron\  
with a template from the Orion Nebula.
The crosses with error bars are the flux ratios corrected  
with the extinction-versus-wavelength relation of Chiar \& Tielens (2006) for the GC;
the asterisks are the same data corrected using the extinction law of Draine (2003a, b).
The solid and dotted lines represent least squares fits through the two sets 
of extinction corrected data, respectively.
(a) The H$_2$ S(0)/S(2) ratio.
The solid line has a slope of $-0.06 \pm 0.02$ and the dotted line 
has a slope of $-0.12 \pm 0.02$.
(b) The H$_2$ S(2)/S(1) ratio.
The solid line has a slope of $-0.001 \pm 0.003$ and the dotted line 
has a slope of $0.195 \pm 0.003$.
}
\end{figure}

\begin{figure}
\epsscale{1.0}
\plotone{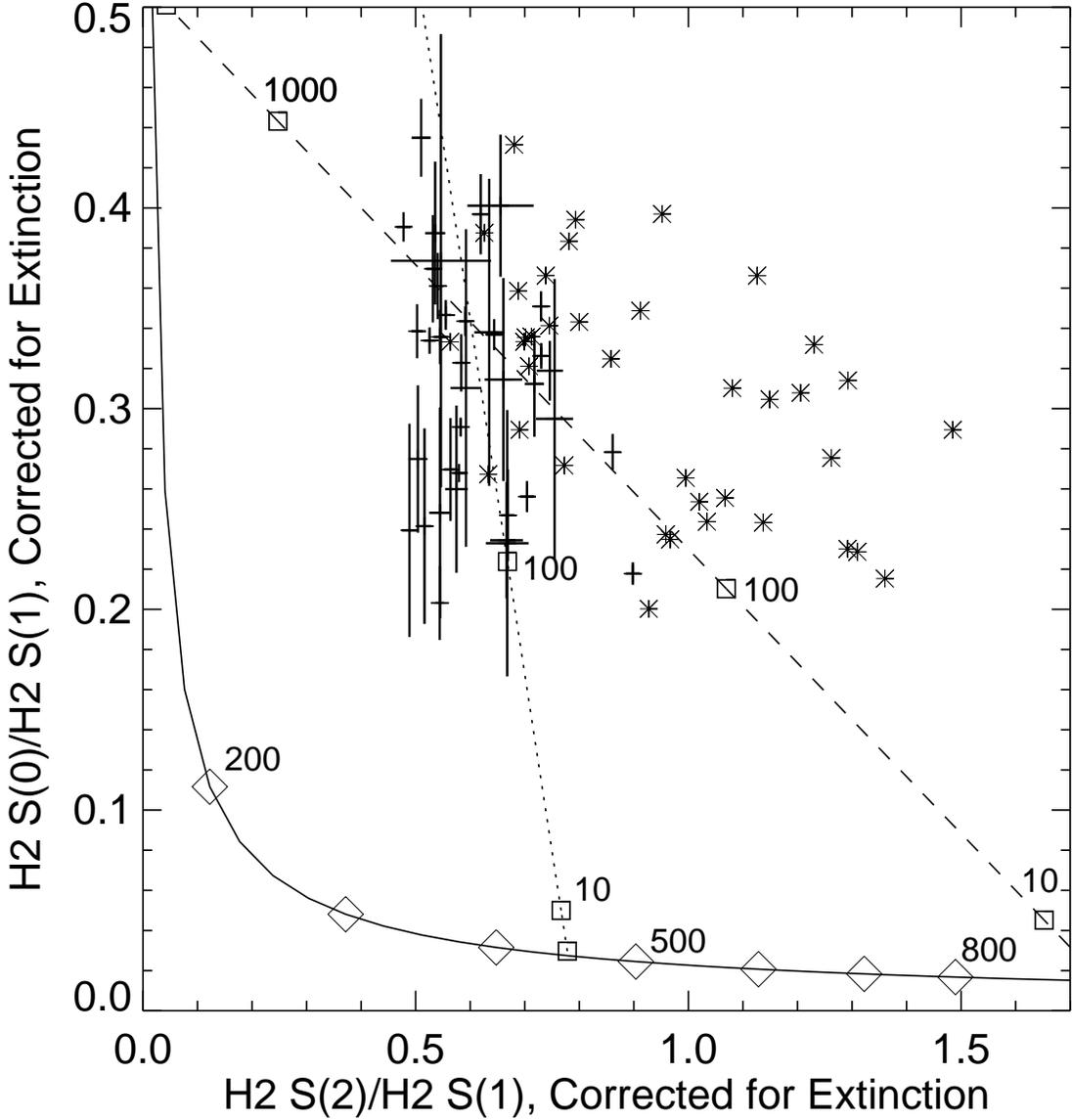}
\caption{The H$_2$ S(0)/S(1) flux ratio plotted versus the 
H$_2$ S(2)/S(1) flux ratio.
The crosses with error bars are the ratios corrected for extinction 
using the extinction law of Chiar \& Tielens (2006) for the GC;
the asterisks are the same data corrected using the extinction law of Draine (2003a, b).
The solid line connects the ratios of the theoretical fluxes of these lines in LTE;
the diamonds mark the gas temperatures, $T$, along the line.
The dashed line shows the locus of ratios of the sum of a cool component with $T = 125$ K 
and a hot component with $T = 1000$ K
and the dotted line shows the locus of ratios of the sum of a cool component with $T = 100$ K 
and a hot component with $T = 450$ K.
The squares mark where the ratio of the column density of the cool component to 
the column density of the hot component is 1000, 100, 10, and 1.
}
\end{figure}

\begin{figure}
\epsscale{1.0}
\plotone{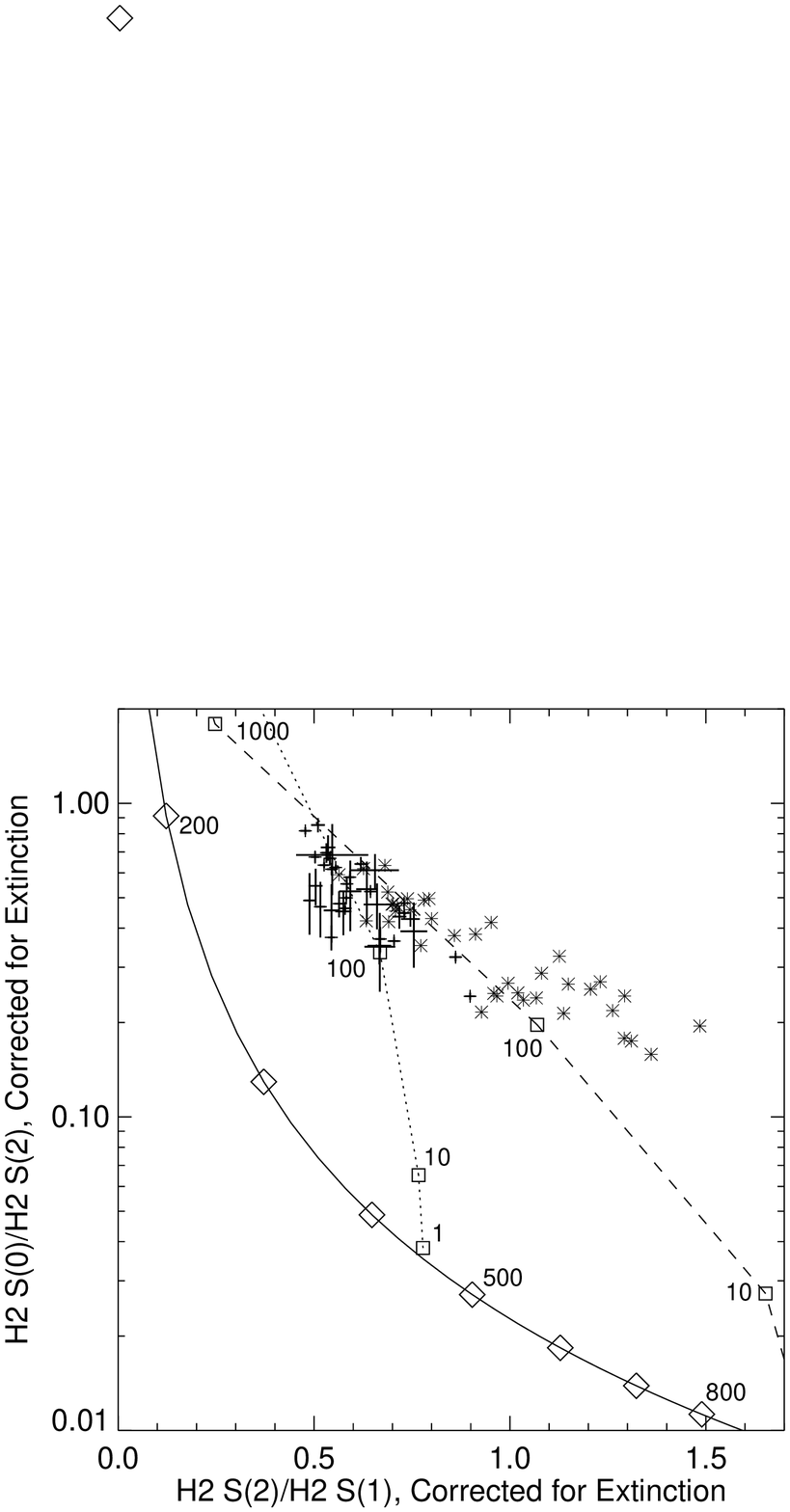}
\caption{The H$_2$ S(0)/S(2) flux ratio plotted versus the 
H$_2$ S(2)/S(1) flux ratio.
See Fig.~16 for a description of the lines and labels.
}
\end{figure}

\end{document}

%% file: tab1.tex

\begin{deluxetable}{lcccccccc}
\tabletypesize{\scriptsize}
\tablecolumns{9}
\tablewidth{0pc}
\tablecaption{Observed Positions}
\tablehead{
\colhead{Position} & \colhead{RA} & \colhead{Dec} & \colhead{Galactic} & 
\colhead{Galactic} & \colhead{$\tau_{9.6}$} & \colhead{$V_{\rm LSR}$} & \colhead{Emission Measure} & \colhead{Notes} \\
\colhead{} & \colhead{(J2000)} & \colhead{(J2000)} &\colhead{Longitude} & \colhead{Latitude} 
& \colhead{} & \colhead{(km s$^{-1}$)} & \colhead{pc cm$^{-6}$} & \colhead{} } 
 
\startdata
\phn 1 &  17 47 05.0 & $-$28 59 12  & $0.1233$ & $-0.2998$ & $0.48$\tablenotemark{a} & $ 10.5$\phn\phn & 1.4E4 & Diffuse ISM \\
\phn 2 &  17 46 58.3 & $-$28 59 45  & $0.1028$ & $-0.2837$ & $1.27$\tablenotemark{a} & $ 10.3$\phn\phn & 1.8E4 & Diffuse ISM \\
\phn 3 &  17 46 53.4 & $-$28 58 32  & $0.1109$ & $-0.2579$ & $1.09$\tablenotemark{a} & $ 15.6$\phn\phn & 2.0E4 &  \\
\phn 4 &  17 46 49.0 & $-$28 58 10  & $0.1078$ & $-0.2410$ & $0.91$ & $  8.5$\phn\phn & 5.1E4 &  \\
\phn 5 &  17 46 45.3 & $-$28 57 45  & $0.1067$ & $-0.2259$ & $1.13$\tablenotemark{a} & $ -0.2$\phn\phn & 4.2E4 &  \\
\phn 6 &  17 46 44.5 & $-$28 57 07  & $0.1142$ & $-0.2179$ & $1.57$\tablenotemark{a} & $ -7.5$\phn\phn & 4.2E4 &  \\
\phn 7 &  17 46 42.5 & $-$28 57 01  & $0.1119$ & $-0.2108$ & $1.17$\tablenotemark{a} & $-9.9$\phn\phn & 5.0E4 &  \\
\phn 8 &  17 46 42.4 & $-$28 55 45  & $0.1297$ & $-0.1996$ & $1.36$\tablenotemark{a} & $-20.9$\phn\phn & 5.1E4 &  \\
\phn 9 &  17 46 36.1 & $-$28 55 46  & $0.1175$ & $-0.1801$ & $1.16$ & $-11.4$\phn\phn & 4.6E4 &  \\
10 &  17 46 35.4 & $-$28 55 15  & $0.1236$ & $-0.1734$ & $1.77$ & $ 26.2$\phn\phn & 8.8E4 & Bubble Rim  \\
11 &  17 46 33.7 & $-$28 55 26  & $0.1177$ & $-0.1697$ & $1.76$ & $ 22.3$\phn\phn & 9.1E4 & Bubble Rim  \\
12 &  17 46 33.2 & $-$28 55 10  & $0.1206$ & $-0.1658$ & $1.55$ & $ 26.1$\phn\phn & 4.6E4 & Bubble Rim  \\
13 &  17 46 30.3 & $-$28 56 16  & $0.0994$ & $-0.1663$ & $2.31$ & $ 4.0$\phn\phn & 1.4E5 & H$_2$O Maser\tablenotemark{b} \\
14 &  17 46 29.9 & $-$28 54 40  & $0.1215$ & $-0.1512$ & $1.72$\tablenotemark{a} & $ 0.7$\phn\phn & 1.9E4 &  \\
15 &  17 46 25.8 & $-$28 54 15  & $0.1196$ & $-0.1348$ & $0.76$ & $-20.5$\phn\phn & 1.6E4 &  \\
16 &  17 46 21.7 & $-$28 53 47  & $0.1185$ & $-0.1180$ & $1.05$ & $ 0.4$\phn\phn & 2.6E4 &  \\
17 &  17 46 23.0 & $-$28 53 08  & $0.1302$ & $-0.1164$ & $1.95$ & $ 18.1$\phn\phn & 4.3E4 &  \\
18 &  17 46 19.1 & $-$28 52 47  & $0.1278$ & $-0.1013$ & $1.74$ & $ 31.3$\phn\phn & 4.2E4 &  \\
19 &  17 46 13.6 & $-$28 51 53  & $0.1302$ & $-0.0763$ & $1.80$ & $-2.0$\phn\phn & 4.2E4 &  \\
20 &  17 46 10.9 & $-$28 52 08  & $0.1215$ & $-0.0701$ & $2.59$ & $-112.6$\phn\phn & 1.6E5 &  \\
21 &  17 46 08.2 & $-$28 51 45  & $0.1218$ & $-0.0583$ & $2.44$ & $-17.7$\phn\phn & 1.0E5 &  \\
22 &  17 46 06.3 & $-$28 50 50  & $0.1313$ & $-0.0445$ & $3.17$ & $ 41.7$\phn\phn & 2.9E5 & Sickle Handle \\
23 &  17 46 02.6 & $-$28 50 52  & $0.1238$ & $-0.0332$ & $2.50$ & $ 6.1$\phn\phn & 1.4E5 &  \\
24 &  17 45 59.3 & $-$28 51 23  & $0.1102$ & $-0.0274$ & $3.11$ & $-7.4$\phn\phn & 2.3E5 &  \\
25 &  17 46 00.5 & $-$28 50 29  & $0.1253$ & $-0.0233$ & $2.91$ & $35.3$\phn\phn & 2.7E5 & Sickle Handle \\
26 &  17 45 58.5 & $-$28 50 40  & $0.1189$ & $-0.0187$ & $2.67$ & $ 5.7$\phn\phn & 1.4E5 &  \\
27 &  17 45 53.9 & $-$28 49 28  & $0.1272$ & \phs $0.0060$ & $2.91$ & $-12.8$\phn\phn & 3.3E5 &  \\
28 &  17 45 50.5 & $-$28 49 20  & $0.1227$ & \phs $0.0179$ & $3.18$ & $-18.4$\phn\phn & 3.8E5 & Arches Cluster \\
29 &  17 45 47.4 & $-$28 48 35  & $0.1274$ & \phs $0.0340$ & $3.11$ & $-13.2$\phn\phn & 2.3E5 &  \\
30 &  17 45 46.5 & $-$28 47 59  & $0.1342$ & \phs $0.0420$ & $3.12$ & $-19.3$\phn\phn & 3.3E5 &  \\
31 &  17 45 45.0 & $-$28 47 49  & $0.1337$ & \phs $0.0481$ & $3.40$ & $-15.0$\phn\phn & 7.2E5 & E2 Filament \\
32 &  17 45 40.7 & $-$28 48 04  & $0.1219$ & \phs $0.0595$ & $2.51$ & $-19.9$\phn\phn & 1.7E5 &  \\
33 &  17 45 35.5 & $-$28 47 25  & $0.1214$ & \phs $0.0812$ & $3.21$ & $-18.3$\phn\phn & 2.5E5 & W Filament \\
34 &  17 45 34.0 & $-$28 47 15  & $0.1209$ & \phs $0.0873$ & $3.02$ & $-21.7$\phn\phn & 5.0E5 & W Filament \\
35 &  17 45 31.9 & $-$28 46 50  & $0.1228$ & \phs $0.0975$ & $2.00$ & $-13.8$\phn\phn & 2.3E5 & W Filament \\
36 &  17 45 21.8 & $-$28 45 26  & $0.1235$ & \phs $0.1411$ & $0.80$ & $-6.9$\phn\phn & 1.1E4 & Diffuse ISM  \\
37 &  17 45 10.5 & $-$28 44 04  & $0.1214$ & \phs $0.1882$ & $0.42$ & $-7.5$\phn\phn & 1.2E4 & Diffuse ISM  \\
38 &  17 44 58.8 & $-$28 42 56  & $0.1152$ & \phs $0.2345$ & $0.33$ & $-7.0$\phn\phn & 9.6E3 & Diffuse ISM  \\
\enddata
\tablenotetext{a}{Positions whose values of $\tau_{9.6}$ were increased 
to produce the minimum [S III] 18.7/33.5 \micron\ line ratio
after correction for extinction.}
\tablenotetext{b}{This maser source (G\"usten \& Downes 1983) has the radio appearance of an Ultra-Compact H II region. Our position is slightly offset from the radio source to avoid foreground stars.}
\end{deluxetable}


%% file: tab2.tex

\begin{deluxetable}{lccccl}
\tablecolumns{6}
\tablewidth{0pc}
\tablenum{2}
\tablecaption{Line Parameters}
\tablehead{
\colhead{Line} & \colhead{Wavelength} & \colhead{$IP$\tablenotemark{a}} & \colhead{$E_{upper}$} & \colhead{$\tau_\lambda/\tau_{9.6}$} & \colhead{References} \\
\colhead{} & \colhead{(\micron)} & \colhead{(eV)} &\colhead{(cm$^{-1}$)} & \colhead{} & \colhead{} }
\startdata
\ H$_2$ S(0) & 28.219  & 0 & 354.37 & 0.369 & 17, 18 \\ 
\ H$_2$ S(1) & 17.035  & 0 & 705.69 & 0.502 & 17, 18 \\ 
\ H$_2$ S(2) & 12.279  & 0 & 1168.80 & 0.344 & 17, 18 \\ 
\ \ion{H}{1} 7-6    & 12.372  & 13.60 & 107440.45 & 0.332 & 13 \\ 
\ [\ion{O}{4}] $^2$P$_{3/2}-^2$P$_{1/2}$   & 25.890 & 54.94 & 386.245  & 0.404 & 4,15 \\ 
\ [\ion{Ne}{2}] $^2$P$_{1/2}-^2$P$_{3/2}$  & 12.814 & 21.56 & 780.42 & 0.313 & 12 \\ 
\ [\ion{Ne}{3}] $^3$P$_1-^3$P$_2$       & 15.555 & 40.96 & 642.88 & 0.383 & 4, 6 \\ 
\ [\ion{Si}{2}] $^2$P$_{3/2}-^2$P$_{1/2}$  & 34.815 & 8.15 & 287.23 & 0.306 & 2, 4 \\ 
\ [\ion{P}{3}] $^2$P$_{3/2}-^2$P$_{1/2}$  & 17.885 & 19.77 & 559.13 & 0.539 & 11 \\ 
\ [\ion{S}{3}] $^3$P$_1-^3$P$_0$         & 33.481 & 23.34 & 298.68 & 0.311 & 4, 16 \\ 
\ [\ion{S}{3}] $^3$P$_2-^3$P$_1$         & 18.713 & 23.34 & 833.06 & 0.548 & 5, 16\\ 
\ [\ion{S}{4}]  $^2$P$_{3/2}-^2$P$_{1/2}$  & 10.511 & 34.79 & 951.43 & 0.777 & 11 \\ 
\ [\ion{Cl}{2}] $^3$P$_1-^3$P$_2$         & 14.369 & 12.97 & 696.00 & 0.310 & 14 \\ 
\ [\ion{Fe}{2}] $^6$D$_{7/2}-^6$D$_{9/2}$ & 25.988 & 7.90 & 384.79 & 0.403 & 1, 7, 9, 10 \\ 
\ [\ion{Fe}{2}] $^6$D$_{5/2}-^6$D$_{7/2}$  & 35.349 & 7.90 & 667.68 & 0.305 & 1, 7, 9, 10 \\ 
\ [\ion{Fe}{2}] $^4$F$_{7/2}-^4$F$_{9/2}$  & 17.936 & 7.90 & 2430.10 & 0.540 & 1, 7, 9, 10 \\ 
\ [\ion{Fe}{2}] $^4$F$_{5/2}-^4$F$_{7/2}$  & 24.519 & 7.90 & 2837.95 & 0.425 & 1, 7, 9, 10 \\ 
\ [\ion{Fe}{3}] $^5$D$_3-^5$D$_4$       &  22.925 & 16.19 & 436.21 & 0.448  & 3, 8, 19 \\ 
\ [\ion{Fe}{3}] $^5$D$_2-^5$D$_3$       &  33.038 & 16.19 & 738.88 & 0.314 & 4, 8, 19 \\ 
\enddata
\tablenotetext{a}{Ionization Potential ($IP$) is the energy required
to produce the ground state of the molecule or ion.}
\tablerefs{
(1) Bautista \& Pradhan 1996;
(2) Dufton \& Kingston 1991;
(3) Erickson et al. 1989;
(4) Feuchtgruber et al. 1997;
(5) Kelly \& Lacy 1995;
(6) McLaughlin \& Bell 2000;
(7) Nussbaumer \& Storey 1988;
(8) Quinet 1996;
(9) Quinet et al. 1996;
(10) Ramsbottom et al. 2005;
(11) Saraph \& Storey 1999;
(12) Saraph \& Tully 1994;
(13) Storey \& Hummer 1995
(14) Tayal 2004;
(15) Tayal 2006;
(16) Tayal \& Gupta 1999;
(17) Turner et al. 1977;
(18) Wolniewicz et al. 1998;
(19) Zhang 1996.
}
\end{deluxetable}

%% file: tab3.tex
\begin{deluxetable}{lccccc}
\tablecolumns{6}
\tablewidth{0pc}
\tablenum{3}
\tablecaption{Observed Line Fluxes\tablenotemark{a}}
\tablehead{\colhead{Position} & \colhead{Galactic} & \colhead{Galactic} & \colhead{Flux(H$_2$ S(0))\tablenotemark{b}} & \colhead{Error(H$_2$ S(0))\tablenotemark{b}} & \colhead{Flux(H$_2$ S(1)), etc.\tablenotemark{b}}  \\
\colhead{} & \colhead{Longitude} & \colhead{Latitude} &\colhead{(W m$^{-2}$)} & \colhead{(W m$^{-2}$)} & \colhead{(W m$^{-2}$)} }
\startdata
1 &  0.1233 & $-$0.2998 & 2.076e-16 & 3.707e-18 & 5.787e-16 \\
2 &  0.1028 & $-$0.2837 & 2.246e-16 & 2.661e-18 & 4.854e-16 \\
\enddata
\tablenotetext{a}{Table 3 is published in its entirety in the
electronic edition of the {\it Astrophysical Journal}.  
A portion is shown here for guidance regarding its form and content.}
\tablenotetext{b}{The electronic edition of this table contains 
the observed fluxes and errors per LH beam area of 
the H$_2$ S(0), S(1), and S(2) 28.92, 17.04, and 12.28 \micron\ lines,
the H 7 -- 6 12.37 \micron\ line, 
the [\ion{O}{4}] 25.89 \micron\ line, 
the [\ion{Ne}{2}] 12.81 \micron\ line, 
the [\ion{Ne}{3}] 15.56 \micron\ line, 
the [\ion{Si}{2}] 34.82 \micron\ line, 
the [\ion{P}{3}] 17.89 \micron\ line, 
the [\ion{S}{3}] 33.48 and 18.71 \micron\ lines, 
the [\ion{S}{4}] 10.51 \micron\ line, 
the [\ion{Fe}{2}] 25.99 and 17.94 \micron\ lines, 
and the [\ion{Fe}{3}] 22.93 \micron\ line.}
\end{deluxetable}

%% file: tab4.tex
\begin{deluxetable}{lcccccc}
\tabletypesize{\scriptsize}
\tablecolumns{7}
\tablewidth{0pc}
\tablenum{4}
\tablecaption{Observed Continuum Fluxes\tablenotemark{a}}
\tablehead{\colhead{Position} & \colhead{Galactic} & \colhead{Galactic} & \colhead{Flux({10.0-10.48 \micron})\tablenotemark{b}} & \colhead{Flux({13.5-14.3 \micron})\tablenotemark{b}}& \colhead{Flux({15.56 \micron})\tablenotemark{b}} & \colhead{Error({15.56 \micron}), etc.\tablenotemark{b}}  \\
\colhead{} & \colhead{Longitude} & \colhead{Latitude} & \colhead{(Jy)} & \colhead{(Jy)} & \colhead{(Jy)} & \colhead{(Jy)} }
\startdata
1 & 0.1233 & $-$0.2998 & 2.32E-01 & 6.39E-01 & 5.36E-01 & 5.60E-03 \\ 
2 & 0.1028 & $-$0.2837 & 2.11E-01 & 5.86E-01 & 4.81E-01 & 4.85E-03 \\
\enddata
\tablenotetext{a}{Table 4 is published in its entirety in the
electronic edition of the {\it Astrophysical Journal}.  
A portion is shown here for guidance regarding its form and content.}
\tablenotetext{b}{The electronic edition of this table contains 
the continuum fluxes per LH beam area 
averaged over the wavelength ranges 10.0 -- 10.48 \micron\ 
and 13.5 -- 14.3 \micron\ and 
the continuum fluxes and errors per LH beam area observed at 
15.56, 18.71, 22.93, 28.22, and 33.48 \micron.}
\end{deluxetable}